\documentclass[12pt]{article}
\usepackage{array}
 \usepackage{amsmath,amssymb}
 \usepackage{amsthm}
 \usepackage{bm}
 \usepackage{latexsym}
 \usepackage{CJK}
 \usepackage{multirow}
 \usepackage{caption}
 \usepackage{graphicx,color}
 \usepackage{amssymb,amsfonts}
 \usepackage{hyperref}
 \usepackage[english]{babel}
 \usepackage{times}
 \usepackage[T1]{fontenc}
 \usepackage{enumerate}
\usepackage{geometry}
\usepackage{float}
\usepackage{appendix}
\usepackage{booktabs}
\usepackage{endnotes}
\usepackage{mathtools}
\usepackage{amscd}
\usepackage{bbm}
\usepackage{multirow}
\usepackage{booktabs}
\usepackage{dsfont}
\usepackage{graphicx}
\usepackage{amsmath}

\usepackage[scaled]{beramono}
\usepackage[T1]{fontenc}
\usepackage[numbers]{natbib}
\usepackage{amsmath,amsfonts,amssymb}
\usepackage{mathtools}
\usepackage{changepage}
\usepackage{bbm}

\usepackage{url}
\usepackage{hyperref}
\usepackage{setspace}
\usepackage{bbm}
\usepackage{fixltx2e}
\usepackage[dvipsnames]{xcolor}
\DeclareMathOperator*{\argmin}{arg\,min}

\newtheorem*{assump_1}{Assumption 1 (Mediator-sharing confounder)}
\newtheorem*{assump_2}{Assumption 2 (Overlap)}
\newtheorem*{assump_3}{Assumption 3 (Consistent surrogate confounder)}
\newtheorem*{assump_4}{Assumption 4 (Null mediator)}
\newtheorem*{the_1}{Theorem 5.1}
\newtheorem*{coro_1}{Corollary 5.1}
\newtheorem{theorem}{Theorem}[section]

\newtheorem{lemma}[theorem]{Lemma}

\usepackage{subcaption}

\newlength{\bibitemsep}
\newlength{\bibparskip}
\let\oldthebibliography\thebibliography
\renewcommand\thebibliography[1]{%
  \oldthebibliography{#1}%
  \setlength{\parskip}{\bibitemsep}%
  \setlength{\itemsep}{\bibparskip}%
}

\begin{document}

\title{{De-confounding causal inference using latent multiple-mediator pathways}}
\author{Yubai Yuan and Annie Qu\footnote{{Yubai Yuan is Assistant Professor, Department of Statistics, The Pennsylvania State University (E-mail: yvy5509@psu.edu). Annie Qu is Chancellor’s Professor, Department of Statistics, University of California, Irvine (E-mail: aqu2@uci.edu). This work is supported by NSF Grants DMS 1952406 and DMS 2210640.}}}
		\date{ }
		\maketitle
		\vspace{-6mm}
		\begin{abstract}

Causal effect estimation from observational data is one of the essential problems in causal inference. However, most estimation methods rely on the strong assumption that all confounders are observed, which is impractical and untestable in the real world. We develop a mediation analysis framework inferring the latent confounder for debiasing both direct and indirect causal effects. Specifically, we introduce generalized structural equation modeling that incorporates structured latent factors to improve the goodness-of-fit of the model to observed data, and deconfound the mediators and outcome simultaneously. One major advantage of the proposed framework is that it utilizes the causal pathway structure from cause to outcome via multiple mediators to debias the causal effect without requiring external information on latent confounders. In addition, the proposed framework is flexible in terms of integrating powerful nonparametric prediction algorithms while retaining interpretable mediation effects. In theory, we establish the identification of both causal and mediation effects based on the proposed deconfounding method. Numerical experiments on both simulation settings and a normative aging study indicate that the proposed approach reduces the estimation bias of both causal and mediation effects. 

				\noindent\textbf{Key words:} Causal identification, Generalized additive model, Latent factor modeling, Mediation analysis, Sequential ignorability 
		\end{abstract}

\section{Introduction}

\begin{figure}[H]
       				\begin{center}
      					\includegraphics[width=3.5in,height=1.3in]{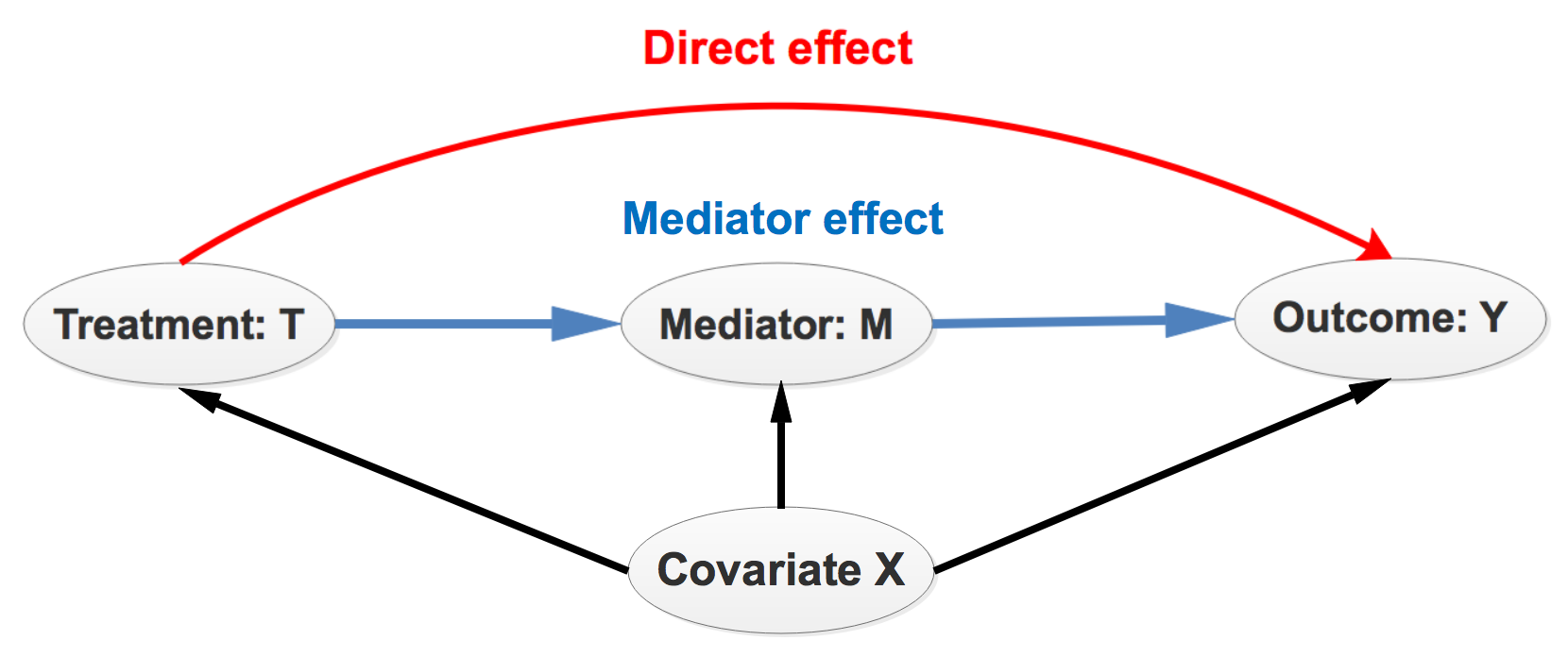}
      				
      				\end{center}
      				\vspace{-5mm}
      				\caption{The causal mediation pathway with a single mediator and observed confounder $X$.}
      				\label{fig_1} 
    \end{figure}

{
Causal inference is one of the most essential tasks for much scientific research to infer whether certain predictors, i.e., treatments have causal effects on outcomes. For complex studies, it is more critical to further identify the mechanism that explains \textit{how} the treatment affects the outcome. One important direction is to investigate how the treatment and outcome relation is transmitted through intermediate variables. Specifically, causal mediation analysis identifies the causal mechanism by delineating pathways from treatment to outcome via mediators \citep{imai2010general}. The basic paradigm of a classic mediation analysis is illustrated in Figure \ref{fig_1}, which is widely used in psychological, sociological, epidemiological and biological studies \citep{hicks2011causal}. }


{In various domain applications, the treatment-outcome mechanism is often complicated and might not be fully captured by a single-mediator model. For example, in study of educational of prevention strategies reducing students' drug addiction, the causal effect of education is explained by various mediators such as resistance skills, social norms, attitudes about drugs, and communication skills \citep{mackinnon2012introduction}. Therefore, the multiple-mediator analysis is often more useful as the causal effects can be decomposed to a number of different mediation pathways, and provides a more accurate assessment and more meaningful interpretation of mediation effects \citep{shi2021testing, cai2020anoce}.}

{The main challenges in both causal inference and mediation analysis are the rise of confounders which could be intervene between treatments and outcomes. {Specifically, confounders introduce non-causal associations between treatments and outcomes, therefore potentially inducing bias in causal-related inference. To solve the confounding issue, most of the existing causal inference methods assume that all confounders are observed. Under this assumption of unconfoundedness \citep{imbens2015causal}, both the causal and mediation effects can be estimated unbiasedly via adjusting the observed data on confounders. For example, adjustment can be achieved via} linear regression \citep{zhang2016estimating, baron1986moderator}, random forest \citep{wager2018estimation}, or other supervised learning methods \citep{kunzel2019metalearners}. However, the assumption of unconfoundedness might not be satisfied in practice, and could also be difficult to verify.} 

{To relax the unconfoundedness assumption, extensive methods have been developed to allow causal identification given the existence of latent confounders. For example,  
\citep{spirtes2013causal, wang2021learning, ogarrio2016hybrid} study conditions where the existence of causal paths from one variable to another can be identified without adjusting the latent confounding effects. However, these methods cannot remove estimation bias on causal effects \citep{spirtes2013causal,ogarrio2016hybrid}, and further require the confounding effect is either significantly stronger or weaker than the causal effect \citep{wang2021learning}. \citep{dukes2021proximal,miao2018identifying, louizos2017causal, tan2006regression} focus on a series of proximal causal inference methods assuming that proxies of the underlying confounders are observed, and the proxy variables are utilized to reduce confounding bias in observational studies.} 

{Recently, a new direction of deconfounding methodology has been developed to adjust the latent confounders via utilizing structural information from specific causal pathways and confounding effects. Specifically, \citep{ranganath2018multiple,wang2019blessings} establish a two-stage deconfounding algorithm utilizing the structure of multiple treatments sharing the same confounder. Similarly, \citep{zhou2020promises} proposes to leverage the structure of multiple outcomes for identification and estimation of causal effects regardless of latent confounders. In addition, \citep{witty2020causal} proposes to reduce the estimation bias of causal effects via incorporating the hierarchical structure of confounding effects where confounders are shared across subjects from the same subgroup. On the other hand, although mediation analysis is important and widely used in real applications, few papers discuss adjusting the latent confounding to reduce estimation bias.}

To address this issue, we develop a mediation analysis framework which allows the identification of causal effects and mediation effects through the latent confounders. Compared with the treatment-outcome pathway considered in  \cite{ranganath2018multiple,wang2019blessings,zhou2020promises}, adjusting latent confounders within causal mediation pathways could be more complex and challenging due to that the {confounder-mediator relations introduce an additional layer of confounding between treatments and outcomes. On the other hand, the mediators themselves contain the latent confounders' information, and can be utilized in a principled way to reduce the biases in estimating causal effects and mediation effects. This motivates us to develop a new deconfounding framework. Specifically, we develop a confounder-sharing structure among multi-mediators, and utilize latent variables to aggregate the information of latent confounders from the shared variations in multiple mediators and outcomes. In addition, the confounder-sharing structure also allows the conditional independence among multiple mediators, which enables us to further estimate the confounder-relevant latent variable through a series of latent modelings. One key innovation is that the proposed method identifies causal and mediation effects without recovering true latent confounders. Instead, the deconfounding latent variables capture the confounder information, and also serve as surrogate confounders to adjust confounding effects.} 

{One advantage of the proposed framework is that it does not require external proxies for latent confounders as in \citep{dukes2021proximal,miao2018identifying, louizos2017causal}. In addition, our method is not restricted to any specific structure in latent confounders themselves, such as the hierarchical structure in \citep{witty2020causal}, and does not impose assumptions on the distribution of latent confounders. 
{Furthermore, the proposed deconfounding algorithm can integrates various nonparametric estimators to infer underlying complex confounding effects on multiple mediators and outcomes, while still retaining the interpretability of the mediation effects.} In theory, we show that the causal effects estimation is unbiased using the proposed surrogate confounders. We also establish the causality identification conditions for our method under different confounder-mediator structures of causal mediation pathways. Numerically, both simulations and the real data application indicate the effectiveness of the proposed method in  reducing estimation biases of both causal effects and mediation effects.} 

{This paper is organized as follows: Section 2 introduces the background of the causal mediation analysis. Section 3 introduces
the proposed deconfounding method. Section 4 provides an algorithm and
implementation strategies. Section 5 establishes the theoretical properties of causal identification in the proposed method. Section 6 demonstrates
simulation studies. Section 7 presents an application to a NIH normative aging study. The last section provides conclusions and some further discussion.}

\section{Background and Notation}

Let $\bm{Y} = \{Y_i\}_{i=1}^N$ denote the set of observed outcomes from $N$ subjects, where $Y_i$ is a one-dimensional outcome from the $i$th subject. Denote $\bm{T} = \{T_i\}_{i=1}^N$ as the set of treatment assignments. We observe multiple mediators for each subject $\bm{M}=\{M^{(j)}_{i}\},\; i=1,\cdots,N, j = 1,\cdots,k$, and covariates $\bm{X} = \{X_i\}_{i=1}^N$ where $X_i \in R^{p}$, and $k,\;p$ are the number of mediators and covariates, respectively. 
To formulate the causal mediation inference mathematically, we adopt the potential outcomes framework \citep{rubin2005causal}. Specifically, let $M^{(j)}_{i}(t)$ denote the potential value of the $j$th mediator for the $i$th subject when the subject takes treatment $t\in \{0,1\}$. Similarly, we use $Y_i(t, m)$ to represent the potential outcome for the $i$th subject given that the subject takes treatment $T_i = t$ with mediators $M_i = (M_i^{(1)},\cdots,M_i^{(k)})$ are $(m_1,\cdots,m_k)$. Notice that observed data can be denoted as $\{M_i(T_i),Y_i(T_i,M_i(T_i))\}$. Under the potential outcomes framework with binary treatment, the average total treatment effect can be defined as $\tau= \mathbb{E}\left\{Y_{i}\left(1, M_{i}(1)\right)-Y_{i}\left(0, M_{i}(0)\right)\right\}$. Based on the total effect, we can specify the portion of the treatment effect through mediators, which is referred to as the average treatment mediation effect (ACME); that is,
$\delta(t) = \mathbb{E}\left\{Y_{i}\left(t, M_{i}(1)\right)-Y_{i}\left(t, M_{i}(0)\right)\right\}.$
Accordingly, the direct treatment effects are defined as 
$\zeta(t) = \mathbb{E}\left\{Y_{i}\left(1, M_{i}(t)\right)-Y_{i}\left(0, M_{i}(t)\right)\right\}.$
It can be shown that the average total treatment effect $\tau = \delta(t) + \zeta(1-t),\;\text{for}\; t= 0,1.$

One essential problem of the mediation analysis in Figure \ref{fig_1} is to directly estimate the total treatment effect and mediation effect
from the observed data. 
One well-established sufficient condition is to require that joint distribution of $\{Y(t, m), M^{(1)}(t),\cdots,M^{(k)}(t)\}$ be independent from the distribution of $T$ conditioning on covariates $X$, and the distribution of $Y(t, m)$ be independent with the distribution of $\big\{M^{(1)}(t),\cdots,M^{(k)}(t)\big\}$ conditioning on $(X,T)$ \citep{imai2010identification}. In other words, the above assumptions require that all the confounders 
of both direct and indirect associations between $T$ and $Y$ 
are observed, which can be stringent in real applications of mediation analysis. Therefore, the unmeasured confounder would introduce non-causal association among treatment, mediators, and outcome, leading to biased estimation for the causal treatment effect and mediation effect.

\section{Methodology}

\begin{figure}[H]
       				\begin{center}
      					\includegraphics[width=2.1in,height=1.35in]{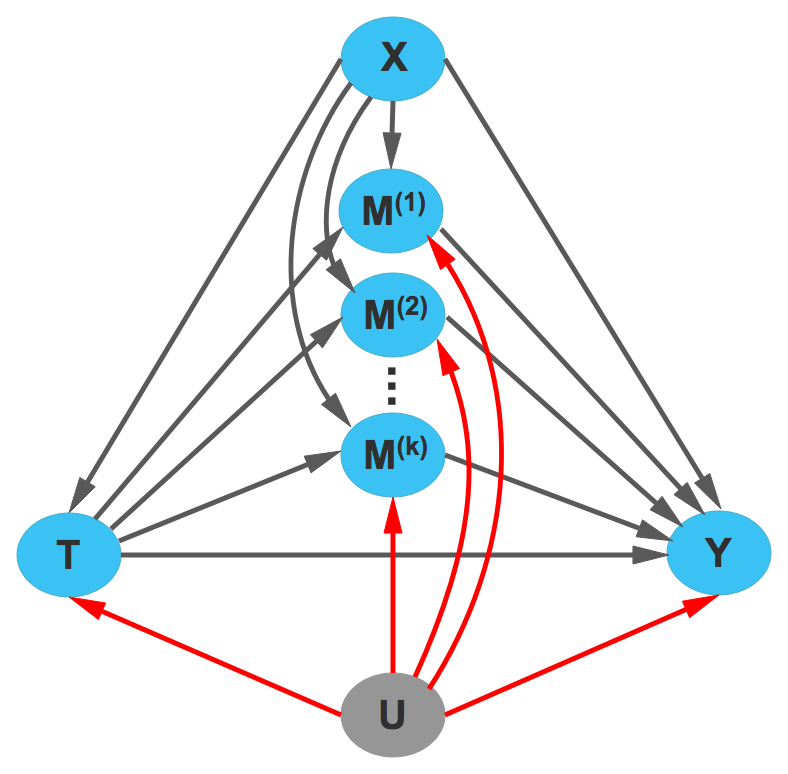}
      				
      				\end{center}
     \vspace{-6mm} 				
      				\caption{The causal mediation pathway with latent confounder $U$ affecting both treatment $T$, outcome $Y$, and multiple mediators $\{M^{(j)}\}_{j=1}^k$. The multiple mediators are not causally dependent on each other. The observed data is colored in blue, and the latent confounder is colored in grey.}
      				\label{fig_2}  
    \end{figure}

In this section, we develop a de-confounder method for a broad class of mediation analysis featuring multiple mediators given the existence of latent confounders. Specifically, we consider the mediation causal pathway where the treatment can have both direct effect and indirect effects via the path of multiple mediators on the outcome, and latent confounders simultaneously affect treatment, outcome, and multiple mediators. The key structure is that the multiple mediators might not be causally dependent on each other. The causal mediation pathway is illustrated in Figure \ref{fig_2}.

%

\vspace{-5mm} 

\subsection{Joint debiasing on multiple-mediator pathway}

{In the following, we formulate the causal mediation pathway in Figure \ref{fig_2} via the potential outcome framework omitting the covariates $X$ for ease of notation. The formulation can be generalized with covariates $X$. We assume that the pretreatment confounder $\bm{U}$ within the causal mediation pathway satisfies the following condition. 
\\
\textbf{Sequential ignorability for multiple mediators:}
\vspace*{-8mm}
\begin{align}
&\left\{Y\left(t^{\prime}, m\right), \bm{M}(t)\right\}  \perp T \mid \bm{U} = u, \label{seq_uncon_1}  \\ 
& Y\left(t^{\prime}, m\right)  \perp \bm{M}(t) \mid T = t, \bm{U} = u,  \label{seq_uncon_2} \\
\hspace{-2.5cm}(\textbf{parallel mediators}):\;  &M^{(i)} \perp \bm{M}^{(-i)}|T = t,  \bm{U} = u,\; i \in \{1,\cdots,k\},
\; t\in \{0,1\}, \label{seq_uncon_3}
\vspace{-6mm}
\end{align}
where $\bm{M}^{(-i)}$ denotes the set of mediators excluding the $i$th mediator. {The conditions (\ref{seq_uncon_1}) and (\ref{seq_uncon_2}), referred to \textbf{sequential ignorability}, are standard assumptions in causal identification for mediation analysis \citep{imai2010identification,forastiere2018principal}.} These two conditions guarantee the identification of mediation effects and direct effects \citep{imai2010identification} via requiring that the joint distribution of mediators and outcome is conditional independent from the treatment assignment, and the distribution of outcomes is conditional independent from the distribution of mediators given the latent confounder $\bm{U}$. {The conditions (\ref{seq_uncon_1})-(\ref{seq_uncon_2}) together imply that $\bm{U}$ is the source of unobserved confounding between multiple mediators and outcomes.}

The condition (\ref{seq_uncon_3}) indicates a parallel-mediator structure in Figure \ref{fig_2} such that the mediators are independent to each other given the latent confounder $\bm{U}$.  The corresponding causal mediation pathway generalizes a broad class of causal inference on observational data in many fields, especially biological and social sciences. For example, it is desirable to uncover how alcohol consumption affects blood pressure via mediators such as body mass index and various enzymes, where certain underlying genes might influence alcohol consumption, blood pressure, and mediators simultaneously \citep{mackinnon2012introduction}. Additionally, it is also of scientific and social interest to investigate how primary prevention programs can reduce drug use via affecting socioeconomic mediators such as resistance skills and social norms. Confounders, such as students' personality, might also influence the causal mediation pathway.

{Since $\bm{U}$ is not directly observed, the conditions (\ref{seq_uncon_1})-(\ref{seq_uncon_3}) are not directly applicable for the identification of mediation effects and direct effects. 
However, the parallel mediator structure allows us to search a surrogate confounder $\bm{\hat{U}} = \{\hat{U}_i\in R^{r}\}_{i=1}^N $ where $r$ is a dimension of the latent vector. With the surrogate confounder, the sequential ignorability of (\ref{seq_uncon_1})-(\ref{seq_uncon_2}) can be approximately established conditioning on $\bm{\hat{U}}$ instead of the unobserved $\bm{U}$.}

{In the following, we introduce a latent factor model based on the surrogate confounder $\hat{\bm{U}}$ to fit the observed data. We denote $\bm{M} = (M^{(1)},\cdots, M^{(k)})$ and the latent factor model as follows:
\begin{align}\label{latent_factor_model_1}
\hat{U}_i \in R^{r} \sim \mathcal{P}(\hat{\bm{U}}),\;
T_i \sim \mathcal{P}(T|\hat{U}_i),\; M_i^{(j)} \sim \mathcal{P}(M^{(j)}|T_i,\hat{U}_i),\; Y_i \sim \mathcal{P}(Y|T_i,\bm{M}_i,\hat{U}_i),
\end{align}
for $i = 1,\cdots, N$ and $j=1,\cdots, k$. The individual latent factors $\{\hat{U}_i\}_{i=1}^N$ introduce the underlying confoundness across $(T,\bm{M},Y)$, which cannot be adjusted from observed data. Assume the latent factor model captures the joint distribution of observed data $\mathcal{P}(T,\bm{M},Y)$ in that
\begin{align}\label{latent_factor_model}
\mathcal{P}(T,\bm{M},Y|\hat{\bm{U}}) =  \mathcal{P}(T|\hat{\bm{U}})\prod_{j=1}^{k}\mathcal{P}(M^{(j)}|T,\hat{\bm{U}}) \mathcal{P}(Y|T,\bm{M},\hat{\bm{U}}).
\end{align} 
Given that the true confounder $\bm{U}$ affects multiple mediators $M^{(1)},\cdots,M^{(k)}$, the decomposition in (\ref{latent_factor_model}) implies the sequential ignorability of (\ref{seq_uncon_1}) and (\ref{seq_uncon_2}) based on $\hat{\bm{U}}$:
\begin{align*}
\left\{Y\left(t^{\prime}, m\right), M^{(j)}(t)\right\}  \perp T \mid \hat{\bm{U}} = u,\;\; 
Y\left(t^{\prime}, m\right)  \perp M^{(j)}(t) \mid T = t, \hat{\bm{U}} = u,  
\end{align*}
indicating that $\hat{\bm{U}}$ can serve as a surrogate confounder. The argument is based on a contradiction from the existence of confoundness  when (\ref{latent_factor_model}) holds and sequential ignorability is  conditioning on $\bm{U}$ in (\ref{seq_uncon_1})-(\ref{seq_uncon_3}). Specifically, if there still exists a latent confounder $\bm{U}$ that simultaneously affects $(T,M^{(1)}(t),\cdots,M^{(k)}(t),Y(t,m))$ and is not captured by $\hat{\bm{U}}$, then the multiple mediators remain dependent even conditioning on $\hat{\bm{U}}$ and $T$ due to the parallel mediator structure (\ref{seq_uncon_3}). Therefore, the decomposition fails in that $\mathcal{P}(\bm{M}|T,\hat{\bm{U}}) \neq \prod_{j=1}^{k}\mathcal{P}(M^{(j)}|T,\hat{\bm{U}})$, and leads to a contradiction in that (\ref{latent_factor_model}) does not hold. In other words, if the latent factor model fits the observed data well and equation (\ref{latent_factor_model}) holds, then we can use $\hat{\bm{U}}$ as the surrogate confounder that contains the information of true confounder $\bm{U}$.} 

{Accordingly, we propose a deconfounding strategy through fitting the observed data of the causal mediation pathway (\ref{latent_factor_model}). We first adopt the additive model to formulate the components $\mathcal{P}(T|{\bm{U}})$, $\mathcal{P}(M^{(j)}|T,{\bm{U}})$, and $\mathcal{P}(Y|T,\bm{M},{\bm{U}})$ incorporating covariate $X$ as follows:
\vspace{-5mm}
\begin{gather}
 Y_i = f_Y(T_i, M_i^{(1)},\cdots,M_i^{(k)}, X_i) + g_Y(U_i) + \epsilon_{Y,i},\nonumber \\ 
 M_i^{(j)} =  f_{M^{(j)}}(T_i, X_i) + g_{M^{(j)}}(U_i) + \epsilon^{(j)}_{M,i},\nonumber \\ 
T_i = f_T(X_i) + g_T(U_i) + \epsilon_{T,i},\; i=1,\cdots,N;\; j = 1,\cdots,k,  \label{pathway_3}
\vspace{-3mm}
\end{gather}
where $f_T(\cdot): R^{p}\to R$, $f_Y(\cdot):R^{k+p+1}\to R$ and $f_{M^{(j)}}(\cdot):R^{p+1}\to R$ are the output functions explaining the treatment assignment $T$, outcome $Y$ and multiple mediators $\{M^{(j)}\}$, respectively. We concatenate the confounding functions $\bm{G}(\cdot): = \{g_T(\cdot),g_{M^{(1)}}(\cdot),\cdots,g_{M^{(k)}}(\cdot),g_Y(\cdot)\}$, which model the effects of individual confounder $U_i$ on each element in $(T_i,M^{(1)}_i,\cdots,M^{(k)}_i,Y_i )$, and  
{therefore $\bm{G}(\cdot): R^r \to R^{k+2}$ encodes the confoundness patterns among the causal mediation pathway through the shared confounder $\bm{U}$.}
In addition, we assume that the permutations $(\epsilon_{Y},\epsilon_{M}^{(1)},\cdots,\epsilon_{M}^{(k)},\epsilon_{T})$ are random variables with zero-mean 
and unknown constant variances. {Notice that the functional forms of  $\{f_T(\cdot),f_Y(\cdot),f_{M^{(j)}}(\cdot),g_Y(\cdot),g_{M^{(j)}}(\cdot),g_T(\cdot)\}$ are not specified. In other words, we do not need to specify explicitly how the underlying true confounder $U$ acts on treatment, mediators, and outcome via $\bm{G}(\cdot)$. Our method works when the model (\ref{pathway_3}) provides a good approximation to the population distribution of observed data.}}

{Based on (\ref{pathway_3}), we can evaluate the fitness of the latent factor model (\ref{latent_factor_model}) for the observed $(T,\bm{M},Y)$ as follows:
\begin{align}\label{req_1}
\vspace{-3mm}
\sum_{i=1}^N\big\{T_i - f_T(X_i) - &g_T(U_i)\big\}^2,\; \sum_{i=1}^N\big\{M^{(j)}_i - f_{M^{(j)}}(T_i,X_i) - g_{M^{(j)}}(U_i)\big\}^2,\nonumber \\ 
&\sum_{i=1}^N\big\{Y_i - f_Y(T_i,\bm{M}_i,X_i) - g_Y(U_i)\big\}^2.
\end{align}
\vspace{-3mm}
{Due to the additive modeling for $\mathcal{P}(T|{\bm{U}})$, $\{\mathcal{P}(M^{(j)}|T,{\bm{U}})\}_{j=1}^k$, and $\mathcal{P}(Y|T,\bm{M},{\bm{U}})$ in (\ref{pathway_3}), the multiplicity in latent factor model (\ref{latent_factor_model}) requires the independence as follows:}
\begin{align}\label{req_2}
\vspace{-3mm}
 \epsilon_T\perp \epsilon_Y, \; \epsilon_T\perp \{\epsilon_M^{(j)}\}_{j=1}^k,\; \epsilon_Y\perp \{\epsilon_M^{(j)}\}_{j=1}^k,\; \epsilon_M^{(i)} \perp \epsilon_M^{(j)},\; i\neq j.
\end{align}
In the following, we denote $\bm{E}_{N\times (k+2)} = (\epsilon_T,\epsilon_M^{(1)},\cdots,\epsilon_M^{(k)},\epsilon_Y)$ and $\bm{F} = \{f_T,f_{M^{(1)}},\cdots,f_{M^{(k)}},f_Y\}$. Combining the observations (\ref{req_1}) and (\ref{req_2}), the proposed framework via the surrogate confounder can be formulated as follows: 
\begin{align}
(\hat{\bm{U}},{\hat{\bm{F}}},\hat{\bm{G}}) = &\argmin_{U,\bm{F},\bm{G}}
\|T - f_T(X) - \bm{G}_1(\bm{U}) \|^2 + \sum_{j=1}^k\|M^{(j)} - f_{M^{(j)}}(T,X) - \!\bm{G}_{j+1}(\bm{U})\|^2  \nonumber \\ 
+&  \|Y- \! f_Y(T,X,\bm{M}) - \bm{G}_{k+2}({\bm{U}})\|^2,\; \text{s.t.}\; \text{corr}(\bm{E})_{(k+2)\times (k+2)} = \bm{I}_{(k+2)\times (k+2)},  
 \label{loss_func}
\end{align}     
where $\text{corr}(\bm{E}) = \big(\text{corr}(\bm{E}_{\cdot i}, \bm{E}_{\cdot j})\big)$ is the correlation matrix of $\bm{E}$. Notice that the residuals $\bm{E}$ can be represented via $(\bm{U},\bm{F},\bm{G})$ as:
\begin{gather*}
\epsilon_T  = T - f_T(X) - g_T({{\bm{U}}}),\; 
\epsilon_Y = Y - f_Y(T,X,{\bm{M}}) - \bm{G}_{k+2}({{\bm{U}}}), \nonumber \\
\epsilon_M^{(j)} = M^{(j)} - f_{M^{(j)}}(T,X) - \bm{G}_{j}({{\bm{U}}}),\; j = 1,\cdots,k.
\end{gather*}    
Therefore, the loss function in (\ref{loss_func}) is optimized jointly over $(\bm{U},\bm{F},\bm{G})$ for the surrogate confounder, output functions, and confounding effect, respectively. The proposed deconfounding strategy (\ref{loss_func}) enables us to construct a surrogate confounder $\hat{\bm{U}}$ and identify the confounding effect $\hat{\bm{G}}(\hat{\bm{U}})$ simultaneously. The former can correct the bias in estimating the causal mediation effect and direct treatment effect, and the latter captures the heterogeneity of mediators and outcomes across different subjects which might not be fully explained by the observed treatments and covariates.}


One advantage of the proposed framework is the flexibility in customizing the functional relations of observed data and latent confounder within outputs to accommodate with different application scenarios. For example, for the parts involving observed data, we can restrict $(f_T, \bm{f}_M, f_Y)$ as parametric model or non-parametric models.
After identifying and separating the confounding effects $\bm{G}(\bm{U})$ from mediators and outcome, the average treatment mediation effect and direct effect can be estimated as          
$\delta(t) = \frac{1}{N}\sum_{i=1}^N \Big[{f}_Y\big\{ t, M^{(1)}(1),\cdots,M^{(k)}(1),X_i \big\} - {f}_Y\big\{t, M^{(1)}(0),\cdots,M^{(k)}(0),X_i \big\}\Big]$, and then
$\zeta(t) = \frac{1}{N}\sum_{i=1}^N \Big[{f}_Y\big\{1, M^{(1)}(t),\cdots,M^{(k)}(t),X_i \big\} - {f}_Y\big\{0, M^{(1)}(t),\cdots,M^{(k)}(t),X_i \big\}\Big].$ Notice that the formation of $\delta(t)$ and $\zeta(t)$ can be further simplified based on specific functional forms of $\{f_M^{(j)}\}$ and $f_Y$.
In the following, we provide detailed discussion on the modeling of latent confounding effects $\bm{G}(\bm{U})$. 

\subsection{Latent confounding effect modeling}

{One key step in the proposed deconfounding strategy is to infer the surrogate confounder $\hat{\bm{U}}$ from the fitted confounding effects $\hat{\bm{G}}(\hat{\bm{U}})$, which does not require the underlying true confounding pattern  $\bm{G}(\cdot)=\big(g_T(\cdot), g_{M^{(1)}}(\cdot),\cdots,g_{M^{(k)}}(\cdot),g_Y(\cdot)\big)$.} In general, we can constrain $\hat{\bm{G}}(\cdot)$ within a class of multivariable functions such that the function class is sufficiently large to approximate $\bm{G}(\bm{U})$. In practice, utilizing the structure in $\bm{G}(\bm{U})$ allows us to determine an appropriate modeling of $\hat{\bm{G}}$.

{Notice that $\bm{G}(\bm{U})$ can be formulated as a $N\times (k+2)$ matrix with $\bm{U} = \{U_i\}_{i=1}^N$.
If the relation among components in $\bm{G}(\cdot)=\big(g_T(\cdot), g_{M^{(1)}}(\cdot),\cdots,g_{M^{(k)}}(\cdot),g_Y(\cdot)\big)$ is governed by linearity, then the $\bm{G}(\bm{U})$ has a low-rank structure in that $\bm{G}_i (\bm{U})$ can be approximated by a linear combination of several $\{\bm{G}_j(\bm{U}),\;j\neq i\}$.} Therefore, the $\bm{G}(\bm{U})$ can be simplified to a latent factor model:
\begin{align}\label{latent_factor}
\bm{G}(\bm{U}) \approx \hat{\bm{G}}(\bm{\hat{U}}) = \hat{\bm{U}}_{N\times r}\bm{A}_{r\times (k+2)},
\end{align} 
where $\bm{A}$ is the loading matrix to be estimated, and $r < k+2$ denotes the rank of the latent confounding matrix. Given that $\hat{\bm{U}}\bm{A}$ captures the subject-wise heterogeneity originating from the variation within $\bm{U}$, then $\bm{\hat{U}}$ would contain the information on the distribution of $\bm{U}$, therefore serving as a confounder surrogate. Here the low-rank structure in $\bm{G}(\bm{U})$ can be verified from the observed data. We fit the mediator and outcome models $(\bm{f}_M,f_Y)$ on the observed data, and then perform the PCA on the residuals $\big( M^{(1)} - \hat{f}_M^{(1)}(T,X),\cdots,M^{(k)} - \hat{f}_M^{(k)}(T,X),Y - \hat{f}_Y(T,X,\bm{M})\big)_{N\times (k+1)}$. If only several leading principle components dominate the variation of residuals, we can choose the latent factor modeling (\ref{latent_factor}) for $\bm{G}(\cdot)$. Accordingly, the rank of surrogate confounder $r$ can be determined via the largest eigengap between two successive eigenvalues.
 
{In many applications, the underlying confounding pattern $\bm{G}(\cdot)$ might be more complex than the linear relations, as the confounding effects $\{g_T(\bm{U}), g_M^{(1)}(\bm{U}),\cdots, g_M^{(k)}(\bm{U}), g_Y(\bm{U})\}$ on treatment, mediators and outcome could be nonlinear to each other. For example, in the causal effect of alcohol consumption on blood pressure, the various enzymes are treated as mediators, and the expression levels of specific genes might confound with the concentration of enzymes while the gene-enzyme relations could be significantly different across different enzymes.} In this case, the latent factor model likely fails as the confounding matrix $\bm{G}(\bm{U})$ is full-rank, and the low-rank structure only preserves the latent space expanded by columns of the nonlinear-transformed $\bm{G}(\bm{U})$.

Alternatively, we adopt the autoencoder \citep{ballard1987modular} which serves as a nonlinear generalization of PCA to infer the information of $\bm{U}$ from $\bm{G}(\bm{U})$. An autoencoder consists of an encoder $\Phi_{encoder}(\cdot)$ and a decoder $\Phi_{decoder}(\cdot)$ where the former performs transformation on the input to extract important features from the observed data and the latter reconstructs the input data based on the extracted features. We estimate the autoencoder via
\begin{align}\label{auto_enc}
(\hat{\Phi}_{encoder},\hat{\Phi}_{decoder}) =  \argmin_{\Phi_{encoder},\Phi_{decoder}} \Big\| \bm{G}(\bm{U}) -  \Phi_{decoder}\Big[\Phi_{encoder}\big\{\bm{G}(\bm{U}) \big\} \Big] \Big\|^2.
\end{align}
Both $\Phi_{encoder}$ and $\Phi_{decoder}$ can be a class of composition function $\{\phi^{(L)} \circ \cdots \phi^{(l)} \circ \cdots \phi^{(1)}:  \phi^{(l)}(\bm{x}) = \varphi(W^{(l)}\bm{x} + b^{(l)})\}$, where $\varphi$ is a nonlinear activation function, $W^{(l)}$ is a weighting matrix, $b^{(l)}$ is a bias vector, and $L$ is the number of layers. With the trained autoencoder $(\hat{\Phi}_{encoder},\hat{\Phi}_{decoder})$, we are able to extract the confounding information within $\bm{U}$ from the encoder $\hat{\bm{U}} = \hat{\Phi}_{encoder}\big\{\bm{G}(\bm{U}) \big\}$. Note that a good fitting of $\bm{G}(\bm{U})$ from (\ref{auto_enc}) indicates $\bm{G}(\bm{U}) \approx \Phi_{decoder}(\hat{\bm{U}})$, which leads to $\hat{\bm{U}} \approx \Phi_{decoder}^{-1}\{\bm{G}(\bm{U})\}$. Here the cardinality of the function space of $\Phi_{decoder}$ increases rapidly as the number of layers $L$ in composition increases. Therefore, $\Phi_{decoder}(\cdot)$ is capable of reconstructing $\bm{G}$ with a higher resolution, in that $\Phi_{decoder}(B) \subset \bm{G}(B)$ for any measurable set $B$ in the input space. Consequently, the $\sigma$-field $\sigma(\bm{\hat{U}})$ is larger than the true confounder $\sigma(\bm{{U}})$, implying that $\hat{\bm{U}}$ can serve as a surrogate confounder to correct all the confounding among treatment assignments, mediators and outcomes to satisfy conditions (\ref{seq_uncon_1}) and (\ref{seq_uncon_2}).

\section{Algorithm and Implementation}

In this section, we develop a deconfounding algorithm for inferring the surrogate confounder via optimizing the proposed objective function (\ref{loss_func}). In general, we sequentially update the output functions $(\hat{f}_T,\hat{\bm{f}}_M, \hat{f}_Y)$, confounding function estimator $\hat{\bm{G}}$ and surrogate confounder $\hat{\bm{U}}$ at each iteration. Different models for the latent confounding effect ${\bm{G}(\bm{U})}$ lead to the different optimization strategies for estimating $\hat{\bm{G}}$ and $\hat{\bm{U}}$. 

For the simplification of presentation, we introduce the $L_{obs}^{(s)}$ to denote the residuals of observed data at the $s$th step as
$$ L_{obs}^{(s)} = \big(T - \hat{f}^{(s)}_T(X), M^{(1)} - {\hat{f}}^{(s)}_{M^{(1)}}(T,X),\cdots, M^{(k)} - {\hat{f}}^{(s)}_{M^{(k)}}(T,X), Y - \hat{f}^{(s)}_Y(T,X,\bm{M}) \big)_{N \times (k+2).}$$
To incorporate orthogonality across the residuals in (\ref{loss_func}), we utilize the method of Lagrange multipliers and transform the constraints into a penalty function as $\|\text{corr}(\bm{E}^{(s)}) - \bm{I}\|_F^2$, where the residuals at the $s$th step are formulated as $\bm{E}^{(s)} = L_{obs}^{(s)} - \hat{\bm{G}}^{(s)}(\hat{\bm{U}}^{(s)})$. 
Therefore, the loss function at the $s$th step is
\begin{align*}
Loss^{(s)} = &\|T - \hat{f}^{(s)}_T(X) - \hat{\bm{G}}^{(s)}_1(\hat{\bm{U}}^{(s)}) \|_2^2 + \sum_{j=1}^k\|M^{(j)} - \hat{f}^{(s)}_{M^{(j)}}(T,X) - \!\hat{\bm{G}}^{(s)}_{j+1}(\hat{\bm{U}}^{(s)})\|_F^2 \\
+ &\|Y- \! \hat{f}^{(s)}_Y(T,X,\bm{M}) - \hat{\bm{G}}^{(s)}_{k+2}(\hat{\bm{U}}^{(s)})\|_2^2 + \lambda \|\text{corr}(\bm{E}^{(s)}) - \bm{I}\|^2_F,
\end{align*}
where $\lambda$ is the Lagrange multiplier. We first illustrate the deconfounding algorithm given that confounding effect $\bm{G}(\bm{U})$ is approximated by the latent factor model, i.e., we replace $\bm{G}(\bm{U}) = \bm{U}_{N\times r} \bm{A}_{r\times (k+2)}$ in (\ref{loss_func}). Notice that with the latent factor modeling, both gradients of the loss function (\ref{loss_func}) in terms of $\bm{A}$ and $\bm{U}$ have explicit forms. 
  
\noindent $\overline{\mbox{\underline{\makebox[\textwidth]{\textbf{Algorithm 1:} Blockwise backfitting for latent factor confounding modeling}}}}$
\begin{enumerate}
\item \textit{(Initialization)} Input the initialization of $(\hat{f}_T^{(0)},\hat{{f}}^{(0)}_{\bm{M}}, \hat{f}_Y^{(0)})$, $\hat{\bm{G}}^{(0)}$, $\hat{\bm{U}}^{(0)}$, Lagrange multiplier $\lambda$, learning rate $\eta$, and stopping threshold $\gamma$.
\vspace*{-3mm}
\item \textit{(Backfitting updates)} At the $s$th iteration $(s\ge1)$.
\vspace*{-3mm}
{\footnotesize \begin{enumerate}
\item[(i)] Given $(\hat{f}_T^{(s-1)},\hat{{f}}^{(s-1)}_{\bm{M}},\hat{f}_Y^{(s-1)})$, $\hat{\bm{A}}^{(s-1)}$, $\hat{\bm{U}}^{(s-1)}$ update the residuals of observed data $L_{obs}^{(s-1)}$. Then update the surrogate confounder via:
\vspace*{-3mm}
\begin{align*}
\hat{\bm{U}}^{(s)} = \hat{\bm{U}}^{(s-1)} - 2\eta \Big\{ \big(L_{obs}^{(s-1)} - \hat{\bm{U}}^{(s-1)}\hat{\bm{A}}^{(s-1)}\big)\hat{\bm{A}}^{(s-1)T} + \lambda\frac{\partial  \|\text{corr}(\bm{E}^{(s-1)}) - \bm{I}\|^2_F}{\partial \bm{U}} \Big\}.
\end{align*}

\vspace*{-5mm}
\item[(ii)] Given $\hat{\bm{U}}^{(s)}$, update the output models $(f_T, f_{\bm{M}},f_Y)$:
\vspace*{-5mm}
\begin{align*}
&\hat{f}^{(s)}_T \leftarrow \argmin_{f_T} \bm{E}_{\hat{\bm{U}}^{(s)}} \big\{\| T -  f_T(X) \|^2| \hat{\bm{U}}^{(s)}  \big\},\\
&{\hat{f}}^{(s)}_{M^{(j)}} \leftarrow \argmin_{f_{M^{(j)}}}\bm{E}_{\hat{\bm{U}}^{(s)}} \big\{ \| M^{(j)} - f_{M^{(j)}}(T,X) \|^2| \hat{\bm{U}}^{(s)} \big\} ,\; j  =1,\cdots,k\\
&\hat{f}^{(s)}_Y \leftarrow \argmin_{f_Y} \bm{E}_{\hat{\bm{U}}^{(s)}} \big\{\| Y -  f_Y(T,X,\bm{M}) \|^2| \hat{\bm{U}}^{(s)}  \big\}.
\end{align*}

\vspace*{-5mm}
\item[(iii)] Given $(\hat{f}_T^{(s)}, \hat{{f}}^{(s)}_{\bm{M}}, \hat{f}_Y^{(s)})$, and $\hat{\bm{U}}^{(s)}$, update the loading matrix $\bm{A}$ via:
\vspace*{-5mm}
\begin{align*}
&\hat{\bm{A}}_{\cdot 1}^{(s)} \leftarrow  \argmin_{v_{r \times 1}}\| T - f^{(s)}_{T}(X) - \bm{\hat{U}}^{(s)}v\|^2,\\
&\hat{\bm{A}}_{\cdot (j+1)}^{(s)} \leftarrow  \argmin_{v_{r \times 1}}\| M^{(j)} - f^{(s)}_{M^{(j)}}(T,X) - \bm{\hat{U}}^{(s)}v\|^2,\; j = 1, \cdots, k, \\
&\hat{\bm{A}}_{\cdot k+2}^{(s)} \leftarrow  \argmin_{v_{r \times 1}}\| Y - f^{(s)}_{Y}(T,X,\bm{M}) - \bm{\hat{U}}^{(s)}v\|^2.
\end{align*}
  
\end{enumerate}}
\vspace{-6mm}
\item \textit{(Stopping Criterion)} Stop backfitting updates if $ \frac{|Loss^{(s)}-Loss^{(s-1)}|}{Loss^{(s-1)}} < \gamma$.
Set ${\hat{f}_{T}} = \hat{f}^{(s)}_{T}$, ${\hat{f}_{\bm{M}}} = {\hat{f}^{(s)}_{\bm{M}}}$, ${\hat{f}_{Y}} = \hat{f}^{(s)}_{Y}$, $\hat{\bm{A}} = \hat{\bm{A}}^{(s)}$, and $\bm{\hat{U}} = \bm{\hat{U}}^{(s)}$. Otherwise set $s \leftarrow s+1$ and iterate Step 2.
\end{enumerate}
\vspace{-0.1in}
\noindent\makebox[\linewidth]{\rule{\textwidth}{0.4pt}}
For the output functions $(f_T,f_{\bm{M}},f_Y)$, we can choose various models such as linear regression, spline, and random forest according to a specific application. For example, when dealing with binary treatment, we adopt the logistic regression model for $f_T$, and change the square loss in (\ref{loss_func}) to a negative log-likelihood loss. In addition, we can initialize $(f_T,f_{\bm{M}},f_Y)$ via fitting each of them on the observed data $(X,T)$, $(T,X, \{M^{(j)}\}_{j=1}^k)$, and $(T,X,\bm{M},Y)$, respectively. Similarly, $\hat{\bm{G}}$ and $\hat{\bm{U}}$ can be initialized via performing the PCA on $(\bm{M},Y)$. Notice that Algorithm 1 can be generalized to the other confounding-effect models where the gradients of the latent confounding effect have explicit forms. In addition, Algorithm 1 can be modified to capture the nonlinear confounding patterns via autoencoder as Section 3.2. The detailed autoencoder-based algorithm is provided in the Section 5 in Supplementary.

\section{Theoretical Results}

{This section establishes the theoretical properties for the proposed deconfouding method. Specifically, we show that the surrogate confounder plays a deconfounding role on treatment assignment, mediators and outcome. In addition, we show that both the causal and mediation effect can be identified by incorporating the surrogate confounders. We also discuss identifiability under different causal pathway structures.} The proposed deconfounding method relies on the structure of confounder-sharing among mediators as follows:  
\begin{assump_1}
There exits a pre-treatment random variable $\bm{U}$ satisfying the following requirements:
\begin{enumerate}
\item Together with $X$ and $T$, $\bm{U}$ generates the smallest $\sigma$-algebra such that individual distributions of mediators are independent from each other
\begin{align}\label{multi_con_1}
M^{(i)}\perp \bm{M}^{(-i)}|T,\bm{U},X. 
\end{align}
\item Together with $X$ and $T$, $\bm{U}$ is the $\sigma$-algebra to satisfy sequential ignorability
\begin{align}\label{multi_con_2}
\left\{Y\left(t^{\prime}, m\right), \bm{M}(t)\right\}  \perp T \mid (\bm{U}, X),\;\text{and}\;
Y\left(t^{\prime}, m\right)  \perp \bm{M}(t) \mid (T ,\bm{U}, X).  
\end{align} 
\end{enumerate}
\end{assump_1}
The mediator-sharing confounder assumption paraphrases the causal mediation pathway in Figure \ref{fig_2}. Due to the conditional independence among mediators, $\bm{U}$ has to contain the information of all the multi-mediator confounders that only affect a subset of mediators $\{M^{(j)}\}_{j\in J},\; J \subset \{1,\cdots, k\},\; 2\leq |J| \leq k$. On the other hand, the concept of smallest $\sigma$-algebra guarantees that $\bm{U}$ only includes confounders affecting multiple mediators and excludes the latent confounders that only affect a single mediator. {The structure of parallel mediators (\ref{multi_con_1}) is common in many scientific and social studies where the multiple mediators are conditionally independent given the treatment and confounders. For instance, in the study of causal relations between mindfulness and emotional distress, the negative cognitive bias and perceived stress are identified as independent mediators \citep{ford2019negative}. In addition, it is shown that interpersonal and intrapersonal factors are parallel independent mediators in the causal effect of HIV stigma on therapy adherence \citep{seghatol2017interpersonal}. Sequential ignorability (\ref{multi_con_2}) is a standard condition in causal mediation inference \citep{imai2010identification}. The first part assumes that the treatment assignment is ignorable given the confounders, which can be satisfied when the treatments are randomly assigned. The second part assumes the ignorability of mediators conditioning on pre-treatment covariates, which is not directly testable from observed data in general \citep{manski2009identification}. However, a set of sensitivity analyses exists to quantify the robustness of causal effect estimation to the potential violation of the ignorability assumption \citep{imai2010general}.}    
\begin{assump_2}
$\mathcal{P}(T = t |\bm{U},X) > 0,\; \text{and} \; \mathcal{P}(\bm{M}=\bm{m}|\bm{U},X) > 0 \; \text{for all $t$ and $\bm{m}$}.$ 
\end{assump_2}
The overlap is a standard condition in the causal inference literature \citep{imai2010general,imai2010identification}, which allows each treatment assignment and mediators' value to have a certain probability to be observed when controlling the confounders. This ensures that the potential outcomes $\{Y(t,m)\}$ can be identifiable. {In the following, we establish the causal identification results based on the surrogate confounder $\hat{\bm{U}}$. Recall that estimating $\hat{\bm{U}}$ via the objective function (\ref{loss_func}) is equivalent to fitting the latent factor model (\ref{latent_factor_model_1}) to the observed data. Specifically, the surrogate $\hat{\bm{U}}$ satisfies 
\vspace{-3mm}
\begin{align}
\mathcal{P}&(T,\bm{M}|X) = \int \prod_{j=1}^k \mathcal{P}(M^{(j)}|T,\hat{\bm{U}},X)\mathcal{P}(T|\hat{\bm{U}},X)\mathcal{P}(\hat{\bm{U}})\; \mathbf{d}\hat{\bm{U}},\label{equ_1}\\
\mathcal{P}(T,\bm{M},Y|X) &= \int  \mathcal{P}(Y|T,\bm{M},\hat{\bm{U}},X) \prod_{j=1}^k  \mathcal{P}(M^{(j)}|T,\hat{\bm{U}},X)\mathcal{P}(T|\hat{\bm{U}},X)\mathcal{P}(\hat{\bm{U}}) \;\mathbf{d}\hat{\bm{U}}, \label{equ_2}
\end{align} 
where $\mathcal{P}(T,{\bm{M}}|X)$ and $\mathcal{P}(T,{\bm{M}},Y|X)$ are the distributions of the observed data while the distributions involving $\hat{\bm{U}}$ under the integrations are inferred from the proposed mediation pathway modeling (\ref{pathway_3}). Intuitively, with the surrogate $\hat{\bm{U}}$, we can decompose the joint distribution of mediators into multiple conditionally independent components to satisfy (\ref{multi_con_1}). Note that there exist different $\hat{\bm{U}}$ compatible with $\mathcal{P}(T,\bm{M}|X)$ and $\mathcal{P}(T,\bm{M},Y|X)$ on the observed data, while the causal identification requires the removal of uncertainty in estimating the latent factor model. Therefore, we allow that $\hat{\bm{U}}$ can be determined by the observed data.  
\begin{assump_3}
All latent variables $\hat{U}$ satisfying  (\ref{multi_con_1}), i.e., $M^{(i)}\perp \bm{M}^{(-i)}\;\big|
\;(T,\hat{U})$, can be consistently identified by the observed treatment and mediators $(T,\bm{M})$ in that for any $\varepsilon>0$,
\vspace{-4mm}
\begin{align}\label{determ}
\vspace{-3mm}
\lim_{n \to \infty} \mathcal{P}(|\hat{U} - f(T,\bm{M})|>\varepsilon) = 0, \; \text{or} \; \lim_{k \to \infty} \mathcal{P}(|\hat{U} - f(T,\bm{M})|>\varepsilon) = 0, 
\vspace{-3mm}
\end{align}
\vspace{-3mm}
where $f(\cdot)$ is a deterministic function dependent on the factor models in (\ref{latent_factor}), $n$ and $k$ are the sample size and number of mediators. 
\end{assump_3}
Notice that this assumption does not require $\hat{\bm{U}}$ to coincide with the true latent confounder $\bm{U}$. Instead, $\hat{\bm{U}}$ only needs to be consistently identified by the observed data. This assumption gives us flexibility in that many choices of low-rank latent models can be used for modeling $\hat{\bm{G}}(\hat{\bm{U}})$ in Section 3.2. For instance, PCA, matrix factorization have the identification property of latent factors when the number of mediators $k$ and the number of samples $N$ are large \citep{chen2020structured, fu2019nonnegative, gresele2020incomplete}. In addition, many practical applications involve a large number of mediators, especially epigenomic studies such as studying the relation between smoking and the risk of lung cancer via high-dimensional DNA methylation markers.

On the other hand, the Assumption 3 can also be satisfied when the number of mediators is finite or small. In this case, we can consistently estimate the surrogate confounder via utilizing prior knowledge about the distribution of confounder $U$. For example, when there exist underlying $K$ subgroups among subjects, we can assume $U_i = a_k\mathbbm{1}_{\{i \in \text{group}\}},\; k\in \{1,2,\cdots,K\}$. Or when mediators change smoothly across similar subjects, one can impose smoothness on the surrogate confounders. The distributional information of $U$ can be incorporated via adding regularizations on the surrogate confounders to encourage a piece-wise constant structure \citep{tang2021individualized}, or control the variations among $U$ from similar subjects \citep{hong2013sparse, wang2017regularized}. Under these constraints, the surrogate confounders can be consistently identified as the number of subjects increases with only limited available mediators or a few of mediators simultaneously affected by $U$. For nonlinear latent factor models, recent developments in identifiability for autoencoder and nonlinear ICA models implies that a surrogate confounder $\hat{U}$ can be consistently estimated by the autoencoder method via good data fitting. Specifically, when the underlying clustering structure exists in $\hat{U}$, then $\hat{U}$ can be consistently identifiable up to the affine transformation \citep{hyvarinen2019nonlinear, 
khemakhem2020variational, willetts2021don}. We perform a theoretical sensitivity analysis for the effects of estimation consistency of the surrogate confounder on causal estimation. See the detailed theorem and discussion of results' in Section 16 of the supplemental material.
\begin{lemma}\label{lemma_1}
Under Assumptions 1-3, the treatment T, mediators $\bm{M}$, and outcome $Y$ are sequential deconfounded given covariate $X$ and surrogate $\hat{\bm{U}}$ satisfying (\ref{equ_1}) and (\ref{equ_2}):
\vspace{-5mm}
\begin{align*}
\vspace{-5mm}
\left\{Y\left(t^{\prime}, m\right), \bm{M}(t)\right\}  \perp T \mid \hat{\bm{U}}, X, \;\;\;
Y\left(t^{\prime}, m\right)  \perp \bm{M}(t) \mid T = t, \hat{\bm{U}}, X,  
\vspace{-5mm}
\end{align*}
\vspace{-3mm}
if the distribution of $(T,\bm{M},Y)$ can be represented as a latent factor model (\ref{latent_factor_model_1}). 
\end{lemma}
{The above Lemma \ref{lemma_1} provides a theoretical justification of the proposed deconfounding scheme in that the surrogate confounder $\hat{\bm{U}}$ contains all the confounding information if the proposed latent modeling captures the distribution of $(T,\bm{M},Y)$. Based on the Lemma \ref{lemma_1}, we establish identification of the average causal effect and mediation effect.}
\vspace{-3mm}
\begin{the_1}\label{theorem_1}
Given that Assumptions 1-3, and the following two conditions holds: (1) the response has additive forms as 
$\mathbb{E}(Y(t,\bm{m})|X=x, \bm{U} = u ) = f_1(t,\bm{m},x)+f_2(u)$ and
$\mathbb{E}(Y|T = t, \bm{M} = \bm{m}, X=x, \hat{\bm{U}} = u ) = f_3(t,\bm{m},x)+f_4(u)$,
where $f_1,f_2,f_3,f_4$ are continuous functions; (2) the mediators follow the generalized additive model in (\ref{pathway_3}).
Then the average mediation effect $\delta(t)$ and average direct treatment effect $\zeta(t)$ can be identified through the surrogate confounder $\hat{U}$ 
\vspace{-4mm}
\begin{align}
\delta(t) = \iint  \Bigg\{ &\int  \mathbb{E}\left(Y\mid T=t,\bm{M}=\bm{m},X = x, \hat{\bm{U}}=u\right) \mathbf{d} F^{(1)}_{\bm{M}}(\bm{m})\nonumber \\ - & \int  \mathbb{E}\left(Y\mid T=t,\bm{M}=\bm{m},X = x, \hat{\bm{U}}=u\right) \mathbf{d} F^{(0)}_{\bm{M}}(\bm{m}) \Bigg\}
 \mathbf{d} F_{X}(x) \mathbf{d} F_{\hat{\bm{U}}}(u), \label{med} \\
\zeta(t) = \iint  \Bigg\{ &\int  \mathbb{E}\left(Y\mid T=1,\bm{M}=\bm{m},X = x, \hat{\bm{U}}=u\right) \mathbf{d} F^{(t)}_{\bm{M}}(\bm{m})\nonumber \\ - & \int  \mathbb{E}\left(Y\mid T=0,\bm{M}=\bm{m},X = x, \hat{\bm{U}}=u\right) \mathbf{d} F^{(t)}_{\bm{M}}(\bm{m}) \Bigg\}
 \mathbf{d} F_{X}(x) \mathbf{d} F_{\hat{\bm{U}}}(u), \label{dir}
 \vspace{-5mm}
\end{align}
where $F_{\bm{M}}^{(t)} = F_{\bm{M} \mid T=t,X = x,\hat{U}=u}(\cdot)$, $F_W(\cdot)$ and $F_{V|W}(\cdot)$ represent the distribution function of a random variable $W$ and the conditional distribution function of $V$ given $W$. Then the average causal effect can be identified as $\tau = \delta(t) + \zeta(1-t),\; t  = 0,1. $
\end{the_1}
\vspace{-3mm}
Theorem 5.1 shows that the surrogate confounder enables the average mediation effect and direct effect to be unbiasedly estimated from the observed data. The result requires additional technical conditions to solve the violation of overlap in that $\mathcal{P}(\bm{M} = \bm{m}|T,X,\hat{\bm{U}})$ might be zero for some $\bm{m}$ due to the deterministic assumption $\hat{\bm{U}}$ in (\ref{determ}).} {However, we show that the identification of the average mediation effect for subsets of the mediators can be achieved under fewer constraints on the distribution of $(T,\bm{M},Y)$. {Notice that the proposed method also applies to the causal pathway where latent confounder only affects multiple mediators, but not affects treatment and outcome. In this case, the surrogate confounder $\hat{U}$ is consistently independent from $T$ and $Y(t,\bm{m})$ since $\hat{U}$ can be consistently captured by the distribution of $\bm{M}(t)$, which is independent to $\{T,Y(t,\bm{m})\}$. Then there does not exist confounding among observed data $(T,\bm{M},Y)$ conditioning on $\hat{U}$.} 
\vspace{-3mm}
\begin{coro_1}\label{cor_1}
Given Assumptions 1-3, the average mediation effect for subsets of mediators $\bm{M}^{(J)} = \{M^{(j)}\}_{j\in J},J\subset \{1,\cdots,k\},|J|<k $ denoted as $\displaystyle \delta^{(J)}(t) = \mathbb{E}\{ Y(t,\bm{M}^{(J)}(1),\bm{M}^{(-J)})\} - \mathbb{E}\{Y(t,\bm{M}^{(J)}(0),\bm{M}^{(-J)})\}$ can be identifiable as
\vspace{-3mm}
\begin{align*} 
\delta^{(J)}(t) = \iint  \Bigg\{ &\int  \mathbb{E}\left(Y\mid T=t,\bm{M}=\bm{m},X = x, \hat{\bm{U}}=u\right) \mathbf{d} F^{(1)}_{\bm{M}^{(J)}}(\bm{m})\nonumber \\ - & \int  \mathbb{E}\left(Y\mid T=t,\bm{M}=\bm{m},X = x, \hat{\bm{U}}=u\right) \mathbf{d} F^{(0)}_{\bm{M}^{(J)}}(\bm{m}) \Bigg\}\mathbf{d} F_{X}(x) \mathbf{d} F_{\hat{\bm{U}}}(u),
\vspace{-5mm}
\end{align*}
where $\bm{M}^{(-J)} = \{M^{(j)}\}_{j\in J^c}$ denotes the complement set of mediators and the conditional distribution $F_{\bm{M}^{(J)}}^{(t)} = F_{\bm{M}^{(J)} \mid T=t,X = x,\hat{U}=u}(\cdot)$. The identification holds when $\bm{M}^{(J)}$ satisfies the overlap condition in that $\mathcal{P}(\bm{M}^{(J)}=\bm{m}|T,X,\hat{\bm{U}})>0$ for any $\bm{m}$.
\vspace{-3mm}
\end{coro_1}
Corollary 5.1 shows that we can estimate the average mediation effect of a subset of mediators unbiasedly utilizing surrogate confounder $\hat{\bm{U}}$ based on observed data. Specifically, it enables us to estimate the mediation effect for each individual mediator. Compared with Theorem 5.1, the identification can be established under weaker assumptions such that the regularity assumptions on the outcome model and mediators are replaced by the overlap condition on subsets of mediators. Notice that the overlap condition on the subset of mediators is compatible with Assumption 3 of the deterministic surrogate confounder in (\ref{determ}), since the consistency $\mathcal{P}(\hat{\bm{U}}|T,\bm{M},Y)=\delta_{f(T,\bm{M}(T))}$ only imposes constraint on the non-zero support of $\mathcal{P}(\bm{M}|T,X,\hat{\bm{U}})$ while the non-zero support of a subset $\bm{M}^{(J)}$ can still be unconstrained.}

{The above identification results are established under Assumption 3 of the deterministic surrogate confounder, which might not be satisfied when the number of mediators is not large enough to exclude the uncertainty in inferring $\hat{\bm{U}}$ from observed data. In the following, we allow an alternative pathway structure for identifying causal mediation and direct effects without Assumption 3. We first introduce the concept of \textit{null mediator} which indicates a mediator $M$ that does not affect the outcome such that $Y\perp M | T,\bm{U},X$. Null mediators are widespread in many scientific studies of  
delineating sparse causal pathways through high-dimensional mediators, such as genes and brain neuroimaging \citep{zhao2016pathway, zhao2020sparse, huang2019genome}. Given a large set of mediators,  only a small subset has nonzero mediation effects between treatments and outcomes. In the following, we introduce the alternative identification condition: 
\vspace{-3.5mm}
\begin{assump_4}
There exists more than one null mediator which do not have causal effect on outcome $Y$, denoted as $\bm{M}^{null}=\{M^{(j)}_{null}: Y \perp M^{(j)}_{null} | T,\bm{U},X \}$ and $|\bm{M}^{null}| \geq 2$. And the null mediators are conditional independent to other non-null mediators $\{M^{(i)}\}_{i=1}^k$ as
\vspace{-4mm}
\begin{align*}
M^{(i)}\perp M^{(j)}_{null}\; \big| \;T,\bm{U},X, \; i = 1,\cdots,k, j = 1, \cdots, |\bm{M}^{null}|.
\end{align*}
\end{assump_4}
\vspace{-3mm}
The null mediators $M^{(k-1)}$ and $M^{(k)}$ are observable descendants of the latent confounder, and therefore can serve as the proxy variables for $\bm{U}$. Notice that the null mediator condition does not require the information as to which mediators are null. Given the existence of null mediators, we can establish the identification of average causal effect as follows. Notice that we do not require the proportion of null mediators for the whole mediator set. To validate the conditional independence among mediators in Assumption 3 and Assumption 4, we can first obtain the surrogate confounder $\hat{U}$ and then adopt the conditional independent test as in \citep{zhang2012kernel, cai2022distribution, huang2010testing, su2008nonparametric}. 
\vspace{-3mm}
\begin{theorem}\label{theorem_2}
Given Assumption 1, Assumption 2, Assumption 4 and weak regularity conditions, the surrogate confounder $\hat{U}$ satisfying (\ref{equ_1}) and (\ref{equ_2}) identifies the average mediation effect $\delta(t)$ as (\ref{med}) and average direct treatment effect $\zeta(t)$ as (\ref{dir}). And the causal effect can be identified as $\tau = \delta(t) + \zeta(1-t),\; t = 0,1$.
\end{theorem}
\vspace{-3mm}
Compared with Theorem 5.1, Theorem \ref{theorem_2} shows that with null mediators, identifying causal mediation and direct effect is still possible even when the surrogate confounder $\hat{\bm{U}}$ is a random variable of the observed data. In another words, it is unnecessary for $\hat{\bm{U}}$ to be completely identified by the observed data to play the deconfounding role. The identification in  Theorem \ref{theorem_2} leverages the proxy variable strategy for direct causal effect \citep{miao2018identifying, kuroki2014measurement} where the distribution of debiased causal effect is identifiable when two proxies of the latent confounder are observed. In the case of multiple mediators, the null mediators can serve as proxies for the shared latent confounder ${\bm{U}}$ to identify other causal mediation pathways. Therefore, unlike the case of no mediator or a single mediator, we do not need to observe external proxy variables for $\bm{U}$. {Notice that Theorem 5.2 still holds when the null mediators are correlated to each other conditioning on unobserved confounder $U$ and treatment $T$ in Assumption 4.}}

\vspace{-4mm}
\section{Numerical Study}
\vspace{-3mm}
In this section, we conduct simulations to investigate the performance of the proposed deconfounding algorithm on debiasing treatment effect estimation, and perform numerical comparisons with existing causal inference and mediation analysis methods. Specifically, our method is compared with existing mediation analysis and causal inference methods: linear structural equation modeling (LSEM), high-dimensional mediation analysis (HIMA), random-forests based causal effect inference (Causal Forest), and meta-learning method (XLearner). These four methods are popular and widely used in many applications. We investigate the debias performance under the settings of linear and nonlinear confounding effects, which are two representative situations of how the latent confounders affect mediators and outcomes.


\vspace{-3mm}
\subsection{Linear confounding effect}
\vspace{-3mm}
In this subsection, we investigate the performance of causal effect estimation based on different methods when the confounding effects on multiple mediators and outcomes are linear representations of each other. Specifically, we assume the subject-wise latent confounders $\bm{U} = \{U_i\}_{i=1}^N$ are randomly generated from a mixture of Guassian distributions as $U_i \sim \omega * N(-2,1.5) + (1-\omega)*N(2,1.5)$ where $\omega \sim \text{Bern}(1,0.5)$. The covariates $X_{N\times p} = \{X_i\}_{i=1}^N$ are generated from $N(\bm{0}_{1\times p}, \bm{I}_{p\times p})$. With the latent confounders $\bm{U}$ and $X$, the observations of $(T,\bm{M},Y)$ are generated via a series of additive models as follows:
\vspace{-5mm}
\begin{align}
& Y = \alpha^{(Y)} T + X\gamma^{(Y)} + \bm{M}\beta^{(Y)}  + g_Y(\bm{U}) + \epsilon_Y, \nonumber\\
 M^{(j)} = T\beta^{(\bm{M})}_j &+ X\Gamma^{(\bm{M})}_{\cdot j} + g_{M^{(j)}}(\bm{U}) + \epsilon_M^{(j)}, \; 
 T_i \sim \text{Bern}\big(\sigma(0.4*U_i)\big),j = 1,\cdots,k,  \label{section_6_1}
\end{align}
where $\sigma(\cdot)$ is the logistic link function, $(g_Y(\bm{U}), g_{M^{(1)}}(\bm{U}),\cdots, g_{M^{(k)}}(\bm{U})) =  (\eta_0\bm{U},\cdots,\eta_{k}\bm{U})$ imposes linear confounding effects on mediators and outcome. In addition, $\big\{\alpha^{(Y)}\in R,\;\beta^{(\bm{M})}=(\beta^{(\bm{M})}_1,\cdots,\beta^{(\bm{M})}_k)\in R^{k}\big\}$, $\big\{\gamma^{(Y)}\in R^{p},\;\Gamma^{(\bm{M})}\in R^{p\times k}\big\}$, $\beta^{(Y)}\in R^{k}$, and $\eta = (\eta_0,\cdots,\eta_k)\in R^{k+1}$ denotes the coefficients of treatment, covariates, mediators, and latent confounders, respectively. Accordingly, the direct causal effect, causal mediation effect and total causal effect can be formulated as $\alpha^{(Y)}$, $(\beta^{(M)})^T\beta^{(Y)}$, and $\alpha^{(Y)} +  (\beta^{(M)})^T\beta^{(Y)}$. In the following simulations, we set the sample size $N=200, 800, 2000$, and the number of mediators $k = 2, 5$. We set coefficients $\alpha^{(Y)}=1$, $\beta^{(M)} = \bm{1}_{1\times k}$, and $\beta^{(Y)} = 0.5\times \bm{1}_{1\times k}$. 

Given that the proposed method, LSEM, and HIMA directly estimate the coefficients $\alpha^{(Y)}$, $\beta^{(M)}$, and $\beta^{(Y)}$, their estimations of causal effect and mediation effect are represented as $\hat{\alpha}^{(Y)}+(\hat{\beta}^{(M)})^T \hat{\beta}^{(Y)}$ and 
$(\hat{\beta}^{(M)})^T \hat{\beta}^{(Y)}$.  
For non-parametric methods such as Causal Forest and XLearner, we obtain the estimation of causal effect via averaging the estimated individual treatment effects. Given that the Causal Forest and XLearner are not designed to directly utilize information of mediators, we combine the mediators $\bm{M}$ with $X$ such that $(X,\bm{M})$ serve as new covariates for model training. In this way, Causal Forest and XLearner include the same amount of information from input as other competing methods for the causal effect estimation to make a fair comparison. The performance is evaluated by the estimation bias of causal effect (Bias$_{total}$) and mediation effect (Bias$_{med}$). Besides the causal effect estimation, we also investigate the performance of predictions of outcomes based on different methods. See Section 7 in supplemental materials for detailed implementation for prediction procedure. 

The performances of causal effect estimations from different methods are illustrated in Figure \ref{fig_sim_1} and Table \ref{sim:1}, where the proposed method utilizing factor modeling (\ref{latent_factor}) and the autoencoder (\ref{auto_enc}) are denoted as \textbf{Prop FM} and \textbf{Prop AE}, respectively. Figure \ref{fig_sim_1} and Table \ref{sim:1} illustrate that the proposed method can consistently achieve a lower estimation bias of causal effect compared with competing methods. As the number of mediators and sample size increase, the 
proposed method achieves a more significant debias in estimation compared to other methods. In particular, in comparison with the best competing method LSEM, our algorithm reduces the causal effect estimation bias by $36\%$ when $k=2$ and by $50\%$ when $k=5$. In addition, the proposed method reduces the estimation bias by $50\%$ to $89\%$ compared to LSEM when the sample size increases from $N=200$ to $N=2000$. Although other methods incorporate information of multiple mediators into causal effect estimation procedure, they do not utilize the confounder-sharing structure among mediators and outcomes. As a consequence, 
the estimation bias accumulates as the number of mediators and sample size increase. However, the proposed method incorporates the inferred confounder to disentangle the confounding among mediators and outcome, and leads to a lower estimation bias. Notice that for the proposed approach, the debias performance based on factor model is better than autoencoder. This is because the factor model can better capture the linear structure of confounding effects on mediators with a lower model complexity than the autoencoder under the current setting.   
\begin{figure}[h]
  \begin{center}
       \hspace*{-0.6cm} 
      					\includegraphics[width=4.5in,height=2.2in]{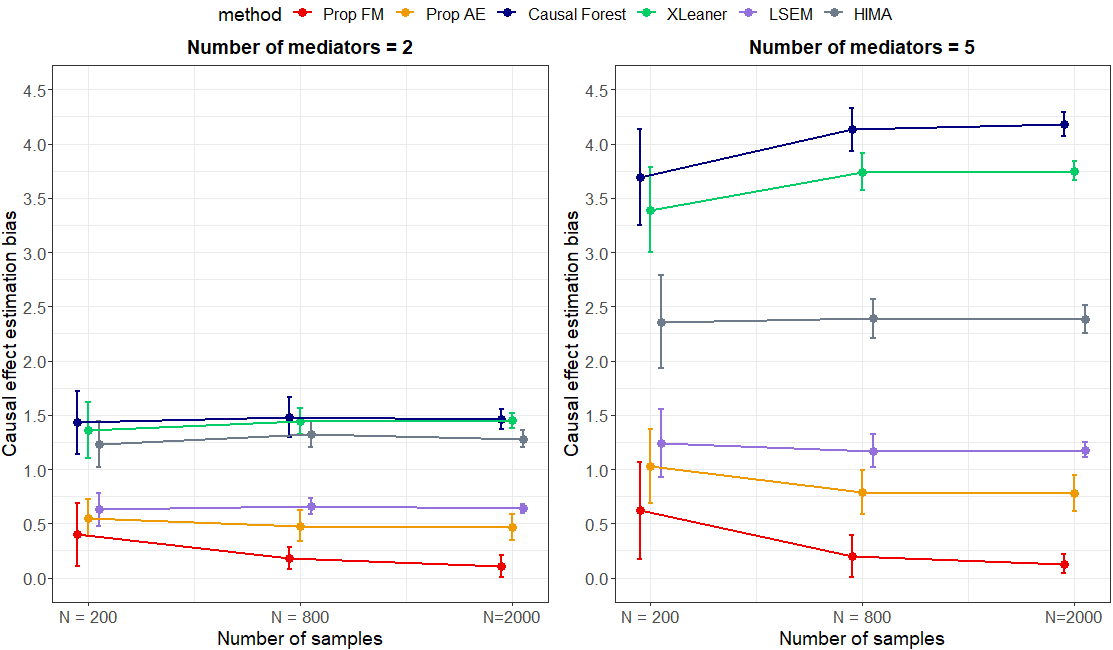}
      				
      				\end{center}
      				\vspace{-5mm}
      				\caption{The bias of causal effect estimation from different methods under the setting of linear confounding effects on multiple mediators and outcomes.}
      	\label{fig_sim_1}  
    \end{figure} 
Given that Causal Forest and XLeaner do not provide the decomposition of causal effect, we compare the proposed method with HIMA and LSEM regarding the mediation effect estimation. The results are illustrated in Table \ref{sim:1}. Similar to the performance in causal effect estimation, the estimation bias of mediation effect from HIMA and LSEM arises significantly as the number of mediators increases while the proposed method retains a low estimation bias. In addition, our method benefits from increasing sample size in terms of reducing bias. Our simulations show the advantage of incorporating a surrogate confounder, which can significantly reduce the confoundness among multiple mediation pathways. 
%
In Table 1, the estimation bias of
the treatment effect increases as the number of mediators increases is due to the estimation bias
of each individual mediator’s effect accumulating when estimating the total treatment effect. However, the average estimation bias $\frac{1}{k+1}|(\hat{\alpha} - \alpha) + \sum_{i=1}^k(\hat{\beta}_i^{M}\hat{\beta}_i^{Y}  - \beta_i^{(M)}\beta_i^{(Y)})|$ decreases as the number of mediators increases.

Furthermore, the performance comparison on outcome prediction is presented in Table \ref{sim:1}. The proposed method consistently achieves a lower prediction error on the testing dataset, and its improvement over the best competing method LSEM increases as the sample size $N$ increases. Table \ref{sim:1} shows that our deconfounding algorithm can utilize multiple mediators to recover partial unobserved effects of latent confounders from outcomes.  
{We conduct simulation studies to investigate robustness of the proposed method when the underlying causal mediation pathway in our setting is misspecified. Specifically, we consider the case when multiple mediators are dependent among each other conditioning on underlying confounder $U$, and the case when $U$ does not affect treatment or outcome. The following simulation results show that our method can achieve lower estimation bias compared with competing methods, and is relatively robust to the misspecification of confounding structure when the sample size is large. The detailed settings and results are provided in Section 8 and 12 of supplemental materials.}    

\begin{table}[h]
\centering
\caption{\small{The performance of causal effect estimation and outcome prediction from different methods under the setting in Section 6.1. The number of mediators are $k=2$ and $k=5$, and the sample size varies from $N=200$ to $N=2000$.}}
\begin{adjustwidth}{1.9cm}{}
\scalebox{0.67}{
\begin{tabular}{| c | c | c | c | c | c | c | c | } 
   \hline\hline

\multirow{3}{*}{Sample size} & \multirow{3}{*}{\textbf{Method}} &   \multicolumn{6}{|c|}{Number of mediators k} \\ \cline{3-8}   
 &   &  \multicolumn{3}{|c|}{k=2} &  \multicolumn{3}{|c|}{k=5}\\ \cline{3-8} 
  &   &  \multicolumn{1}{|c|}{Bias$_{total}$}  & \multicolumn{1}{|c|}{Bias$_{med}$} & MSE &
 \multicolumn{1}{|c|}{Bias$_{total}$} &  \multicolumn{1}{|c|}{Bias$_{med}$}  & MSE 
 \\ \hline

\multirow{5}{*}{$N = 200$}  & \textcolor{red}{Prop AE} & 0.53(0.18) & 0.33(0.21) & 0.89(0.11)  & 1.10(0.27) & 0.80(0.62)  &  0.99(0.10) \\ 

  & \textcolor{red}{Prop FM} & 0.35(0.21)   & 0.26(0.18) & 0.89(0.10) & 0.54(0.48)  & 0.68(0.60) &  1.00(0.10)  \\ \cline{2-8}

 & Causal Forest & 1.41(0.28)  & --  & 1.13(0.15)  & 3.77(0.49) & -- & 1.46(0.12) \\ 

 &XLearner & 1.36(0.26)  & --  & 1.11 (0.13) & 3.33(0.37) & -- & 1.49(0.19) \\ 
 & LSEM & 0.64(0.13)  &  1.14(0.13) & 0.96(0.10)  & 1.19(0.23) & 2.76(0.27) & 1.09(0.11)\\

& HIMA & 1.29(0.24)  &  1.44(0.21) & 1.06(0.13)  & 2.42(0.52) & 3.59(0.47) & 1.21(0.11)
 
  \\ \hline

\multirow{5}{*}{$N = 800$}  & \textcolor{red}{Prop AE} & 0.51(0.12)  & 0.14(0.13) & 0.86(0.04)  & 0.81(0.20) & 0.32(0.19) & 0.96(0.03)  \\
 
  & \textcolor{red}{Prop FM} & 0.18(0.15)   & 0.12(0.08) & 0.86(0.04) &0.21(0.17)&0.26(0.19) & 0.95(0.03) \\ \cline{2-8}

 & Causal Forest & 1.48(0.11)  & -- & 1.00(0.05)  &4.09(0.17) & -- & 1.22(0.06) \\ 

 &XLearner & 1.44(0.12)  & --  & 0.99(0.04) & 3.71(0.14) & -- & 1.24(0.06) \\ 
 & LSEM & 0.66(0.05)   &  1.15(0.06) & 0.93(0.04)  & 1.18(0.14) & 2.73(0.13) & 1.06(0.03) \\
 
& HIMA & 1.26(0.12) &  1.38(0.09) & 1.04(0.07)  & 2.31(0.19) & 3.49(0.19) & 1.15(0.05)
 
 \\ \hline
 
\multirow{5}{*}{$N = 2000$}  & \textcolor{red}{Prop AE} & 0.43(0.10) & 0.13(0.08) & 0.87(0.02)  & 0.75(0.20) & 0.21(0.16) & 0.94(0.04)  \\ 

  & \textcolor{red}{Prop FM} & 0.09(0.09)  & 0.12(0.10) & 0.87(0.03) &  0.12(0.09)& 0.24(0.18) & 0.96(0.04)  \\ \cline{2-8}

 & Causal Forest & 1.46(0.08)  & -- & 0.96(0.03)  &4.16(0.11) & -- & 1.15(0.04) \\ 

 &XLearner & 1.45(0.07)  & -- & 0.96(0.03) & 3.76(0.07) & -- & 1.17(0.05)\\ 
 & LSEM & 0.64(0.04)  &  1.13(0.04) & 0.93(0.03)  & 1.17(0.06)  &2.73(0.07) &1.04(0.03)  \\
 
& HIMA & 1.27(0.07) &  1.42(0.06) & 1.04(0.04)  & 2.34(0.12) & 3.51(0.10) & 1.15(0.04)
 
 \\ \hline

\hline
\hline
\end{tabular}}
\label{sim:1} 
 \end{adjustwidth}
 
\end{table}

\vspace{-3mm}
\subsection{Nonlinear confounding effect}
\vspace{-3mm}
In this subsection, we perform numerical comparisons when the observed data $(T,\bm{M},Y)$ are generated following the sequential models (\ref{section_6_1}) in Section 6.1, and the confounding effects on multiple mediators and outcome $\bm{G}(\bm{U}) = \{g_{M^{(1)}}(\bm{U}),\cdots,g_{M^{(k)}}(\bm{U}),g_{Y}(\bm{U})\}$ are generated as different nonlinear functions of $\bm{U}$. We first consider the low-rank confounding effect as follows:
\vspace*{-.5cm}
\begingroup\makeatletter\def\f@size{10}\check@mathfonts
\begin{align}
&g_{M^{(k)}}(\bm{U}) = \text{Piecewise}(a^{(k)},b^{(k)}),k = 1,2,3,\;
g_{M^{(4)}}(\bm{U}) = \sin(\bm{U}),\nonumber \\[-10pt]
&g_{M^{(5)}}(\bm{U}) =  g_{M^{(1)}}(\bm{U})\times\cos(\bm{U}),
g_{Y}(\bm{U}) = g_{M^{(1)}}(\bm{U})\times \exp(-\bm{U}/6),  \label{section_6_2_1}
\vspace*{-.5cm}
\end{align}
\endgroup
where $\text{Piecewise}(a,b)$ is denoted as the piece-wise function on $\bm{U}$ i.e., and $\displaystyle\text{Piecewise}(a,b) = \sum_{l=1}^{|a|}a_l \mathds{1}_{ \{b_{l} \leq \bm{U} < b_{l+1}\}}$, with $a,\;b$ are the piece-wise function values and cutoffs. We set $a^{(1)} = (1,2,-1,-2,-3)$, $a^{(2)} = (-2,0.5,1,2,3,4)$, $a^{(3)} = (-1,2,3)$; and $b^{(1)} = (-\infty,-3,-1,1,3,\infty)$, $b^{(2)} = (-\infty,-4,-2,0,2,4,\infty)$, $b^{(3)} = (-\infty,-3,3,\infty)$ according to the above data generating process. In addition to the above settings in (\ref{section_6_2_1}), we also investigate other settings of nonlinear confounding effect as follows: 
\vspace*{-.8cm} 
\begingroup\makeatletter\def\f@size{10}\check@mathfonts
\begin{align}
&g_{M^{(1)}}(\bm{U}) = \mathds{1}_{\{T = 0\}}\big\{\text{Piecewise}(a^{(4)},b^{(1)})\big\} + \mathds{1}_{\{T = 0\}}\big\{\text{Piecewise}(a^{(5)},b^{(1)})\big\}, \nonumber \\[-10pt]
&g_{M^{(2)}}(\bm{U}) =  (0.5 + T)\times \text{Piecewise}(a^{(6)},b^{(2)}),  \nonumber  \\[-10pt]
&g_{M^{(3)}}(\bm{U}) = 2\times \mathds{1}_{\{T = 0\}}\big\{\text{Piecewise}(a^{(7)},b^{(3)})\big\} + 2\times \mathds{1}_{\{T = 0\}}\big\{\text{Piecewise}(a^{(8)},b^{(3)})\big\}, \nonumber  \\[-10pt]
&g_{M^{(4)}}(\bm{U}) =  2T\times \big\{\sin(\bm{U})+0.2\big\},\;
g_{M^{(5)}}(\bm{U}) =  g_{M^{(1)}}(\bm{U})\times \big\{\cos(\bm{U})+0.5\big\},\nonumber \\[-10pt]
&g_{Y}(\bm{U}) = \exp(-\bm{U}/8 + 0.9T), \label{section_6_2_2}\vspace{-6mm}
\end{align}
\endgroup
with $a^{(4)} = (1,2,-2,-1,1)$, $a^{(5)} = (2,3,0,-1,2)$, $a^{(6)} = (-2,0.5,1,2,1,-1)$, $a^{(7)} = (-1,0,-1)$, and $a^{(8)} =(-0.5,1-0.5)$. Although both (\ref{section_6_2_1}) and (\ref{section_6_2_2}) are highly nonlinear in terms of $\bm{U}$, the former setting leads to a low-rank structure in $\bm{G}(\bm{U})$ in that 
the first principle component contributes about $55\%$ total variation to $\{\bm{G}({U_i})\}_{i=1}^N$, while the proportion of the first component only explains about $38\%$ of the total variation of $\{\bm{G}({U_i})\}_{i=1}^N$ for the latter setting. In other words, the confounder $\bm{U}$ can be captured via a linear combination of mediator-wise confounding effects for the former setting. In addition, setting (\ref{section_6_2_2}) introduces the interaction between treatment and confounder in the confounding effects to mimic the post-treatment effect, where association between treatment and mediators, and between mediators and response are influenced by the treatment. The sample size varies from $N=200, 800, 2000$ and $N=1000, 2000, 3000$ for settings (\ref{section_6_2_1}) and (\ref{section_6_2_2}), respectively.   
  
{The performance of causal effect estimation and mediation effect estimation are illustrated in Figure \ref{fig_sim_non} and Table \ref{sim_non:1}. Under the low-rank setting, the proposed method with factor modeling  achieves a lower estimation bias, and its improvement over the best competing method LSEM increases as the sample size increases. The improvement patterns are similar to the ones under linear confounding effects in Section 6.1, showing that the proposed factor modeling is able to capture the variation of subject-wise confounders regardless of the functionality of confounding effects under the low-rank structure of $\bm{G}(\bm{U})$. Due to the interactions between $T$ and $U$, the data-generating process under setting (\ref{section_6_2_2}) is in fact misspecified for the deconfounding framework here which is satisfied under a homogeneous treatment effect model (\ref{pathway_3}). Correspondingly, the estimation biases from Prop FM and Prop AE both increase while these biases still remain lower than other competing methods. Our simulation results suggest that the proposed method is robust against model misspecificaiton, and still effective under the heterogeneous treatment effect setting to some extent. Different from the low-rank setting (\ref{section_6_2_1}), the Prop AE produces smaller bias than Prop FM under the setting (\ref{section_6_2_2}) when the relation among confounding effects of mediators and outcome is more complex, and beyond linearity. Through strong representation power from the autoencoder, Prop AE extracts more of the confounding information of $\bm{U}$ shared by $\{g_{M^{(1)}}(\bm{U}),\cdots,g_{M^{(k)}}(\bm{U}),g_{Y}(\bm{U})\}$ compared with factor modeling.}    

{In terms of outcome prediction, Table \ref{sim_non:1} shows that the nonparemetric methods Causal Forest and XLearner provide smaller prediction MSE than the two parametric methods HIMA and LSEM, since the tree-based methods such as Causal Forest and XLearner can better capture the nonlinear confounding effects on outcome. However, the proposed method still outperforms compared to the Causal Forest and XLearner, as we can incorporate the structural information of parallel mediators via the surrogate confounders.}
\begin{figure}[h]
  \begin{center}
       \hspace*{-0.6cm} 
      					\includegraphics[width=4.5in,height=2.2in]{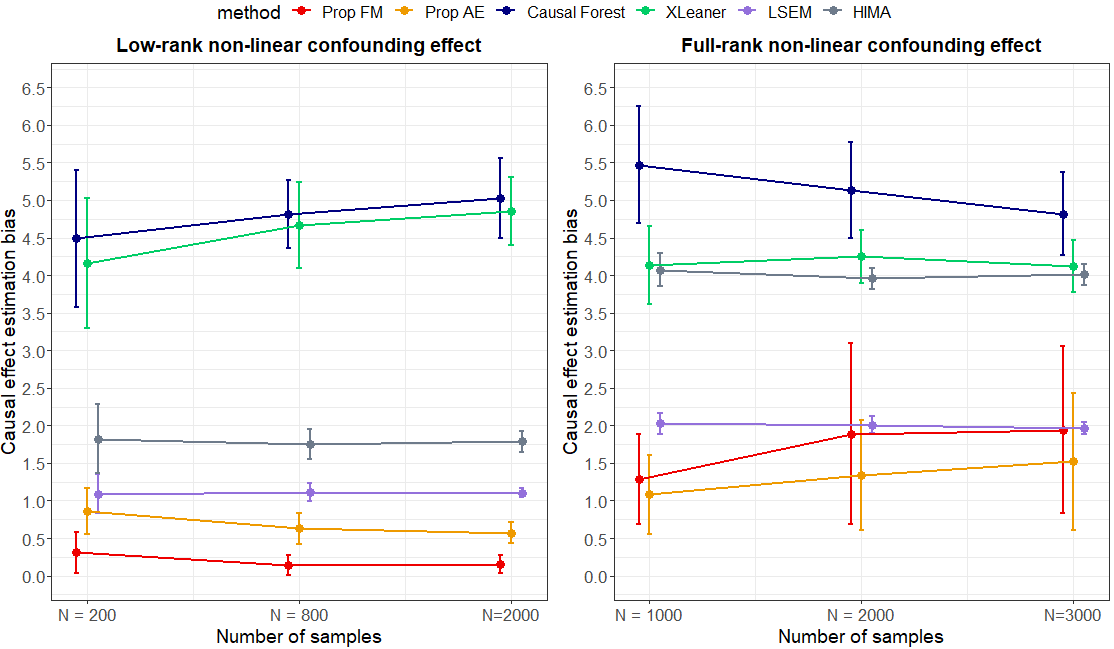}
      				
      				\end{center}
      				\vspace{-5mm}
      				\caption{\small{The bias of causal effect estimation from different methods under the setting of nonlinear confounding effects on multiple mediators and outcome.}}
      	\label{fig_sim_non}  
    \end{figure} 
    \vspace{-5mm}

%
%
%
%

\begin{table}[h]
\centering

\caption{\small{The performance of causal effect estimations and outcome predictions from different methods under the setting in Section 6.2. The number of mediators is $k=5$. }}
\begin{adjustwidth}{1.9cm}{}
\scalebox{0.68}{
\begin{tabular}{| c | c | c | c | c | c | c | c | c | } 
   \hline\hline

 \multirow{2}{*}{\textbf{Method}}  &   \multicolumn{4}{|c|}{Low-rank nonlinear confounding effect } &  \multicolumn{4}{|c|}{Full-rank nonlinear confounding effect }\\ \cline{2-9} 
  & Sample size  & \multicolumn{1}{|c|}{Bias$_{total}$}  & \multicolumn{1}{|c|}{Bias$_{med}$} & MSE & Sample size &
 \multicolumn{1}{|c|}{Bias$_{total}$} &  \multicolumn{1}{|c|}{Bias$_{med}$}  & MSE 
 \\ \hline

 \textcolor{red}{Prop AE} &  \multirow{6}{*}{$N = 200$}  & $0.86_{0.31}$ & $0.94_{0.93}$ & $1.84_{0.56}$  & \multirow{6}{*}{$N = 1000$} & $1.08_{0.52}$ & $0.97_{0.72}$  &  $2.28_{0.11}$ \\ 

   \textcolor{red}{Prop FM} & & $0.31_{0.27}$   & $1.01_{0.78}$ & $1.66_{0.57}$ &  &  $1.29_{0.60}$ & $1.25_{0.43}$ &  $2.23_{0.11}$  \\ 

  Causal Forest & & $4.49_{0.91}$  & --  & $2.23_{0.50}$ &  & $5.47_{0.88}$ & -- & $2.42_{0.08}$ \\ 

  XLearner & &  $4.16_{0.86}$ & --  & $2.19_{0.07}$ &  & $4.14_{0.52}$ & -- & $2.38_{0.09}$ \\ 
  LSEM & & $1.09_{0.26}$  &  $3.55_{0.44}$  & $2.53_{0.66}$ & & $2.03_{0.14}$ & $3.50_{0.23}$ & $2.59_{0.10}$\\

 HIMA & & $1.82_{0.46}$  &  $2.39_{0.86}$ & $2.53_{0.66}$  & &  $4.07_{0.22}$ & $4.58_{0.72}$ & $2.65_{0.18}$
 
  \\ \hline

   \textcolor{red}{Prop AE} &  \multirow{6}{*}{$N = 800$} & $0.63_{0.21}$ & $0.79_{0.51}$ & $1.64_{0.27} $ & \multirow{6}{*}{$N = 2000$} & $1.34_{0.73}$ & $1.37_{0.82}$ & $2.21_{0.09}$  \\
 
   \textcolor{red}{Prop FM} & & $0.14_{0.13}$   & $0.79_{0.56}$ & $1.64_{0.27}$  & & $1.89_{1.20}$& $1.93_{0.86}$ & $2.19_{0.08}$ \\ 

  Causal Forest & &  $4.81_{0.45}$  & -- & $1.92_{0.24}$  & & $5.13_{0.64}$ & -- & $2.32_{0.09}$ \\ 

  XLearner & & $4.67_{0.57}$  & --  & $1.88_{0.22}$ & & $4.25_{0.35}$ & -- & $2.30_{0.08}$ \\ 
  LSEM & &  $1.11_{0.12}$   &  $3.62_{0.18}$ & $2.05_{0.23}$  & & $2.01_{0.11}$ & $3.62_{0.13}$ & $2.56_{0.05}$ \\
 
 HIMA & & $1.75_{0.20}$ &  $2.82_{0.30}$ & $2.19_{0.23}$  & & $3.96_{0.14}$ & $4.48_{0.34}$ & $2.62_{0.09}$
 
 \\ \hline
 
  \textcolor{red}{Prop AE}  & \multirow{6}{*}{$N = 2000$}  & $0.57_{0.14}$ & $0.57_{0.40}$ & $1.44_{0.17}$ & \multirow{6}{*}{$N = 3000$}   & $1.52_{0.91}$ & $1.59_{0.66}$ & $2.21_{0.06}$  \\ 

   \textcolor{red}{Prop FM} & & $0.16_{0.12}$  & $0.47_{0.42}$ & $1.42_{0.17}$ & &  $1.94_{1.11}$& $1.83_{0.86}$ & $2.18_{0.07}$  \\ 

  Causal Forest & & $5.03_{0.53}$  & -- & $1.75_{0.15}$ &  & $4.82_{0.55}$ & -- & $2.28_{0.06}$ \\ 

  XLearner  & & $4.85_{0.45}$  & -- & $1.72_{0.16}$ & & $4.12_{0.35}$ & -- & $2.27_{0.06}$\\ 
  LSEM & &  $1.10_{0.06}$  &  $3.63_{0.13}$ & $2.03_{0.13}$ &  & $1.96_{0.08} $ &$3.54_{0.17}$ & $2.56_{0.06}$  \\
 
 HIMA & & $1.79_{0.14}$ &  $2.77_{0.20}$ & $2.27_{0.17}$ &  & $4.01_{0.14}$ & $4.56_{0.26}$ & $2.63_{0.08}$ 
 \\ \hline

\hline
\hline
\end{tabular}} 
\label{sim_non:1}
 \end{adjustwidth}
 
\end{table}

\vspace{-4mm}
\section{Real Data Example}
\vspace{-3mm}
{In this section, we apply the proposed deconfounding algorithm to the Normative Aging Study data obtained from the NIH dbGaP database (\href{https://www.ncbi.nlm.nih.gov/gap/}{https://www.ncbi.nlm.nih.gov/gap/}) under \href{https://www.ncbi.nlm.nih.gov/projects/gap/cgi-bin/study.cgi?study_id=phs000853.v1.p1&phv=222004&phd=&pha=&pht=4429&phvf=&phdf=&phaf=&phtf=&dssp=1&consent=&temp=1#attribution-section}{phs000853.v1.p1}. The Normative Aging Study (NAS) is a longitudinal study conducted by the United States Department of Veterans Affairs starting from 1963, which collects phenotype data and genotype data from 657 male participants via regular physical examinations and laboratory tests. The phenotype data consists of status of coronary heart disease, diabetes, hypertension, Apolipoprotein E4 protein, and different types of white blood cells from individual blood tests. In addition, the phenotype data include the basic sociodemographic information of smoking status as a binary variable, age at death, and years of education. The genotype data contains the DNA methylation levels at 26,987 individual CpG sites for each participant, which are measured via the Infinium Human Methylation450 technique \citep{dedeurwaerder2014comprehensive}.

{Studies have shown that smoking is hazardous to individual health regarding adverse effects on the quality of life and effects on death risk, and deterioration of the individuals' health status through various pathways in lifestyles and diseases \citep{carbone2005smoking,doll1956lung, peto1994smoking}. In addition, recent studies discovered that smoking causes extensive genome-wide changes in DNA methylation \citep{zeilinger2013tobacco, lee2013cigarette}, which plays a critical role in the development and progression of cancers, and immune-system-related complex diseases \citep{jin2018dna, suarez2012dna}. Motivated by these findings, our goal is to investigate whether the smoking habit further affects the lifespan based on the NAS data, and whether there are causal mediation pathways from smoking to lifespan through DNA methylation. Specifically, we focus on estimating the effect of smoking status on reducing the participants' lifespan, and identify mediation effects by DNA methylation levels at different CpG sites.}

Given that the original DNA methylation data is high-dimensional and contains methylation levels from CpG sites which could be irrelevant to the mediation pathway, we first preprocess data to select significant mediators. Specifically, we select methylation levels at 22 CpG sites serving as mediators in the following analysis. The detailed preprocessing are provided in Section 9 in Supplementary. We investigate the smoking effect on lifespan predictions based on the proposed method and four competing methods: LSEM, HIMA, Causal Forest, and XLearner, as detailed in Section 6. The prediction mean square error is evaluated via 5-fold cross-validation. We provide the implementation of our deconfounding algorithm and other methods in the Supplementary. The results are provided in Table \ref{real_data} where Treat$_{total}$, Treat$_{dir}$, and Treat$_{med}$ denote the estimation of total treatment effect, direct effect, and mediation effect. The prediction error indicates the medians of lifespan prediction mean square errors from multiple repeated measurements.
\begin{table}[h]
\centering
\caption{\small{Estimations of total treatment effect, direct effect, mediation effect, and prediction error of lifespan estimations from different methods for the NAS data where Prop FM and Prop AE stand for the proposed methods modeling confounding effect with factor model and autoencoder, respectively.}} 
\vspace{-2mm}
\resizebox{0.55\textwidth}{!}{\begin{tabular}{| c | c | c | c | c |} 
   \hline\hline
 &  \multicolumn{1}{|c|}{Treat$_{total}$} & \multicolumn{1}{|c|}{Treat$_{dir}$} & \multicolumn{1}{|c|}{Treat$_{med}$} &
Prediction Error   \\ \hline

Prop FM  & -8.40$_{1.01}$  & -1.95$_{0.87}$ & -6.45$_{0.60}$ & 2.94$_{0.27}$  \\ \hline

Prop AE & -8.30$_{1.06}$  & -1.58$_{0.37}$  & -6.72$_{1.09}$ & 3.15$_{0.33}$   \\ \hline \hline

Causal Forest & -1.60$_{0.34}$  & -- & -- & 3.21$_{0.30}$ \\ \hline
XLearner & -1.59 $_{0.22}$  & -- &  -- & 3.13$_{0.27}$   \\ \hline
LSEM & -1.86$_{0.19}$  &   -1.69$_{0.22}$ & -0.16$_{0.22}$ & 3.03$_{0.29}$ 
\\ \hline
HIMA & -1.97$_{0.26}$  &   -1.51$_{0.20}$ & -0.45$_{0.14}$ & 39.57$_{15.58}$ \\ 
\hline 
\hline
\end{tabular}}
\label{real_data}
\end{table}
{Compared with existing methods, the proposed method detects a significantly stronger adverse effect of smoking on lifespan via incorporating the latent confounding effects. Specifically, the proposed method utilizing either latent factor modeling or autoencoder estimates, shows that the smoking habit reduces the lifespan by about 8 years for participants in the NAS study, while estimations from other methods are less than 2 years. On the other hand, a national study based on the 2004 US Census data concludes that the effect of smoking on increasing the mortality risk is similar to reducing 5 to 10 years in lifespan \citep{woloshin2008risk}. In addition, other smoking-cancer association studies show that the average loss of life for smokers is 8 years in Europe and United States \citep{boyle1997cancer}. Therefore, our method produces a treatment effect of smoking on lifespan, which is more consistent with existing data. In addition, clinical studies found that smoking can significantly increase the risk of premature death \citep{jha2008nationally,gavin2004smoking}, which also supported a stronger negative effect of smoking on lifespan. }  

In addition to the direct treatment effect, both the proposed method, LSEM, and HIMA can identify the roles of DNA methylation levels as mediators to transmit the indirect effect of smoking on lifespan, supported by scientific evidence that smoking-induced DNA methylation also increases the risks of metabolic disorders, chronic diseases, diabetes, and cancers \citep{tsai2018smoking, maas2020smoking, besingi2014smoke, jamieson2020smoking}. However, in contrast to the existing LSEM and HIMA, the proposed method can capture stronger mediation effects of DNA methylation in that the proportion of total treatment effect of smoking is almost fully mediated by DNA methylation levels. The DNA methylation level is found to have a full mediation effect of smoking on epigenetic aging \citep{lei2020effect}, $90\%$ on bladder cancer risk \citep{jordahl2019differential}, and $55\%$ on lung functional degradation \citep{de2018blood}. In addition, \citep{zhang2016smoking} found that the DNA methylation level is the most informative biomarker for predicting risks from all causes and cardiovascular mortality associated with smoking \citep{zhang2016smoking}. These studies indicate a significant and dominant role of DNA methylation levels in conveying the smoking effect reduction of lifespan, which is also consistent with the estimation of mediation effects from the proposed method. The distinct gaps between our deconfounding method and existing non-deconfounding methods on both total treatment effect and mediation effects suggest the usefulness of latent confounders for the NAS study. Furthermore, our deconfounding algorithm with latent factor modeling also produces a lower prediction error of lifespan estimation through incorporating partial confounding effects on lifespan. {We also provide the interpretation for the latent confounder inferred from the 22 CpG mediators, and the sensitivity analysis of the causal estimation on the choice of factor models in Section 9 of the supplemental material.}

%
%
%
%
%
%
\vspace{-4mm}

\section{Discussion} 
\vspace{-3mm}
{In this paper, we propose a novel deconfounding method and algorithm to debias causal effect and mediation effect estimation in causal mediation analysis. Specifically, we consider the causal pathways with parallel mediators such that mediators are causally independent conditioning on the shared latent confounders. Our method generalizes the classic mediation analysis paradigm, and is applicable for a wide range of applications. The proposed method utilizes the confounder-mediator structure in multi-channel mediation pathways to infer the information of latent confounders. In addition, we provide flexible modeling on the confounding mechanism regarding the effects of latent confounders from treatment, mediators, and outcomes.} The principle idea of our method is to construct surrogate confounders incorporating the confounding information instead of recovering the original confounders.  In theory, we establish sequential ignorability via incorporating surrogate confounders. Accordingly, we show that both the causal effect and mediation effect can be identified based on the joint distribution of observed data and surrogate confounders. In particular, we provide identification conditions for causal effect estimation under different mediation pathway structures. Our numeric studies also confirm that the proposed method reduces the estimation bias of causal effect and mediation effect under various confounding mechanisms through the confounder-sharing structure of multiple mediators. 
In this paper, the surrogate confounders are jointly estimated. In order to incorporate new observations, we are required to re-train the model on the entire dataset to update surrogate confounders for both  
historical and new observations. The proposed method can be modified using an online learning scheme in that deconfounding for new observations can be computationally independent from historical data. 
\vspace{-3mm}
\section*{Acknowledgements} 
\vspace{-3mm}
The authors thank the Associate Editor and anonymous reviewers for their suggestions and helpful feedback which improved the paper significantly. 
\vspace{-8mm} 
\section*{Supplementary Materials}
\vspace{-3mm}
The supplementary materials provide proofs of the Lemma 5.1,  Theorem 5.1, Corollary 5.1, Theorem 5.2, deconfounding algorithm using autoencoder, illustrations of numerical comparisons in Section 6, and data preprocessing, implementation and interpretation for real data application, 
sensitivity analysis, simulations under misspecified causal pathways, discussion on examples of real applications, and discussion on relation to deconfounder for multiple causes.

\vspace*{-.7cm}

{\footnotesize \bibliography{my_bib, response, response_2}}
\end{document}